%% file: main.tex
\documentclass[10pt,twocolumn,letterpaper]{article}

\usepackage[pagenumbers]{cvpr} 

\usepackage{graphicx}
\usepackage{amsmath} 
\usepackage{amssymb}
\usepackage{booktabs}

\usepackage[pagebackref,breaklinks,colorlinks]{hyperref}
\usepackage{comment}

\usepackage[capitalize]{cleveref}
\crefname{section}{Sec.}{Secs.}
\Crefname{section}{Section}{Sections}
\Crefname{table}{Table}{Tables}
\crefname{table}{Tab.}{Tabs.}

\newcommand\blfootnote[1]{%
  \begingroup
  \renewcommand\thefootnote{}\footnote{#1}%
  \addtocounter{footnote}{-1}%
  \endgroup
}

\def\confYear{2024}

\begin{document}

\title{Sophia-in-Audition: Virtual Production with a Robot Performer}

\author{
Taotao Zhou*$^{1,2}$
\quad
Teng Xu*$^{1,2}$
\quad
Dong Zhang$^{1,2}$
\quad
Yuyang Jiao$^{1}$
\quad
Peijun Xu$^{1}$
\\
Yaoyu He$^{1}$
\quad
Lan Xu$^{1}$
\quad
Jingyi Yu$^{1}$
\\
{$^{1}$ShanghaiTech University}
\quad
{$^{2}$LumiAni Technology}
\\
{\tt\small \{zhoutt2023, xuteng, zhangdong, jiaoyy2022, xupj1, heyy2022, }
\\
{\tt\small xulan1, yujingyi\}@shanghaitech.edu.cn}
}

\twocolumn[{
\renewcommand\twocolumn[1][]{#1}
\maketitle
\begin{center}
    \centering
    \captionsetup{type=figure}
    \includegraphics[width=1.0\textwidth]{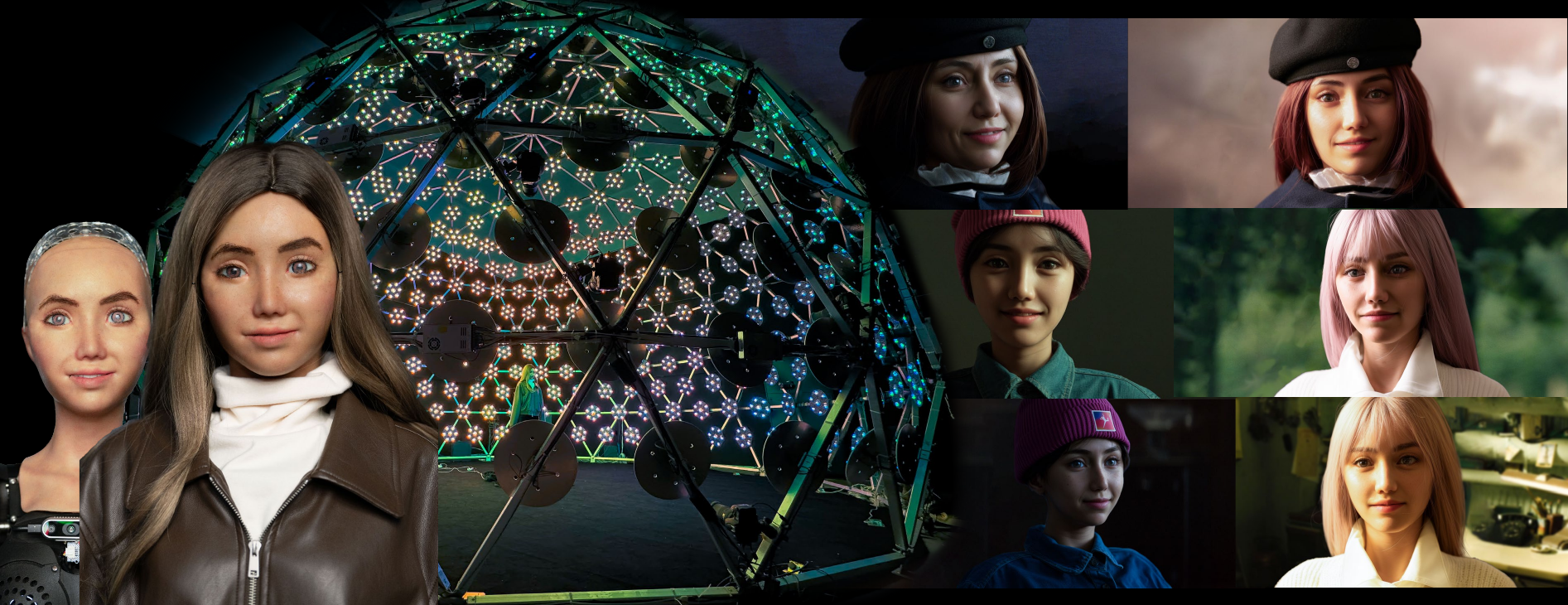}
    \captionof{figure}
    {
    \textbf{Sophia-in-Audition(SiA).} We present a new practice of virtual production: we deploy the humanoid robot Sophia as a virtual performer inside a virtual production studio, in our case, an UltraStage composed of a controllable lighting dome analogous to Light Stage coupled with multi-camera video shooting. We call this setup Sophia-in-Audition or SiA which allows for simultaneous controls over performance, lighting, and camera movements.
    }
    \label{fig:teaser}
\end{center}
}]

\maketitle
\blfootnote{*Authors contributed equally to this work.}

\input{sec/0_abs}

\input{sec/notation}
\input{sec/1_intro}
\input{sec/2_related}
\input{sec/3_system}

\input{sec/4_dataset}

\input{sec/5_idenity}

\input{sec/6_virutal_lighting_and_cammove}

\input{sec/7_userstudy}
\input{sec/8_discussion}

{\small
\bibliographystyle{ieee_fullname}
\bibliography{egbib}
}

\end{document}

%% file: sec/0_abs.tex
We present Sophia-in-Audition (SiA), a new frontier in virtual production, by employing the humanoid robot Sophia within an UltraStage environment composed of a controllable lighting dome coupled with multiple cameras. We demonstrate Sophia's capability to replicate iconic film segments, follow real performers, and perform a variety of motions and expressions, showcasing her versatility as a virtual actor. Key to this process is the integration of facial motion transfer algorithms and the UltraStage's controllable lighting and multi-camera setup, enabling dynamic performances that align with the director's vision. Our comprehensive user studies indicate positive audience reception towards Sophia's performances, highlighting her potential to reduce the uncanny valley effect in virtual acting. Additionally, the immersive lighting in dynamic clips was highly rated for its naturalness and its ability to mirror professional film standards. The paper presents a first-of-its-kind multi-view robot performance video dataset with dynamic lighting, offering valuable insights for future enhancements in humanoid robotic performers and virtual production techniques. This research contributes significantly to the field by presenting a unique virtual production setup, developing tools for sophisticated performance control, and providing a comprehensive dataset and user study analysis for diverse applications.

%% file: sec/notation.tex
\newcommand{\xt}[1]{{\color{red}{[Xuteng: #1]}}}
\newcommand{\ztt}[1]{{\color{magenta}{[Taotao: #1]}}}
\newcommand{\yu}[1]{{\color{red}{[Prof. Yu: #1]}}}
\newcommand{\tcite}[0]{{\color{green}{{[cite] }}}}

%% file: sec/1_intro.tex
\section{Introduction}
\label{sec:1_intro}
A successful shot in filmmaking requires a collaborative effort where performance, lighting, and camera movements are carefully orchestrated to bring the director's vision to life. Performance is the soul of the film to faithfully convey emotions, intentions, and the essence of the character. Lighting and shadows augment the emotional impact, guiding the viewer's attention, and contributing to the overall aesthetic. In fact, many award-winning shots reflect a thorough understanding of the interplay of light and shadow, color temperatures, and the emotional weight that lighting can carry in a scene. Camera movements, in contrast, are a product of extensive practice and understanding of visual language. Movement as simple as a deliberate shooting angle contributes to the narrative. Overall, the process of coordinating all three components - performance, lighting and camera movement - is essential to production but at the same time, they require years of training and are expensive, labor-intensive, and time-consuming.

Over the past decade, significant advances in imaging and cinematographic technologies have led the adoption of virtual production (VP), a more affordable approach for filmmakers to create high-quality content on smaller budgets. On lighting and camera movement, successful examples include the pioneering uses of pre-visualization and virtual camera work to create the illusion of characters floating in space in the feature film \textit{Gravity} \cite{Gravity} and the use of a large LED video wall known as ``The Volume" to produce realistic dynamic virtual environments in Disney's \textit{Mandalorian} series \cite{Mandalorians}. In particular, ``The Volume" manages to reproduce convincing lighting on the characters' faces to match the virtual environments but it is prohibitively expensive. Light Stage \cite{debevec2000acquiring} and its extensions \cite{hawkins2001photometric, debevec2002lighting, wenger2005performance, debevec2012light} instead use a dome with controllable light to capture the performance under controlled variant lighting. They not only serve as a low-cost alternative to the LED wall but also can achieve post-capture re-lighting. Despite these advances in virtual environments, by far nearly all virtual productions require having real actors to perform onsite, even in an audition. In this paper, we explore the possibility of having a robot give an audition performance in a Light Stage setting.

The use of robots is rare, if any, in virtual production. The majority of efforts have been focused on generating lifelike replicas of actors as well as altering scenes post-production. AI-based techniques such as Deepfakes \cite{deepfakes} have made significant strides in creating realistic digital doubles and deepfakes without the need for costly reshoots. Deepfakes, however, generally require experienced performers to start with, before their appearances are re-enacted. They also easily fail in complex lighting conditions. Recently, generation techniques based on Stable Diffusion(SD) \cite{rombach2022high} represent a significant leap in image and even video generation \cite{blattmann2023stable}, supporting high-resolution \cite{wang2022zero, yue2023resshift}, realistic textures \cite{huang2023humannorm}, and even partial lighting effects \cite{ponglertnapakorn2023difareli, yin2023cle}. Yet SD is not yet ready to generate coherent videos, especially for producing accurate facial expressions and motion to reflect subtle emotion. This limitation stems from SD's inherent limitations on maintaining consistency across frames, capturing subtle nuances of motion, and aligning facial expressions over time.

We present a new practice of virtual production: we deploy Sophia, a humanoid robot developed by Hanson Robotics \cite{Sophia}, as a virtual performer inside a virtual production studio, in our case, an UltraStage composed of a controllable lighting dome analogous to Light Stage coupled with multi-camera video shooting. We call this setup Sophia-in-Audition or SiA which allows for simultaneous controls over performance, lighting, and camera movement. Sophia is made from ``Frubber" (flesh rubber), which mimics human musculature and skin, allowing her to simulate more than 60 human-like facial expressions. We demonstrate how to drive Sophia to conduct a variety of performances including replicating classic or iconic segments from movies, following real performers, and performing via AI-generated motion and expressions. To bridge the gap between Sophia's motor-based facial controls with existing CG models (e.g., the BlendShape \cite{blendshape}), we present tailored algorithms to conduct facial motion transfer while maintaining the emotion, critical to any successful performance. 

Having Sophia perform within the UltraStage further allows a user to control environment lighting and camera moves. Our UltraStage contains 480 programmable LED light panels of six spectrums. To reproduce the environment lighting of iconic clips, we first estimate high-dynamic range environment maps from the original clip and set out to optimize the irradiance that best emulates the estimated ones. We show our estimation scheme is effective and manages to replicate many classic settings (Fig.~\ref{fig:dataset}). The UltraStage also uses 32 cameras distributed around the dome. By synchronizing them and shooting Sophia's performance from different perspectives, we can conduct virtual camera moves. Specifically, we extend the latest 3D Gaussian Splatting \cite{kerbl3Dgaussians} to our setting to enable as simple as virtual re-shooting at a new angle and sophisticated camera moves.  

SiA is particularly useful for virtual trial shots. Within this virtual production setup, users can easily experiment with different lighting setups or revisit classic ones and assess various camera movements and framing choices. More importantly, the use of Sophia allows for an effective evaluation of how well the performance matches the lighting and camera moves. It also enables onsite, dynamic performance adjustment, to ensure the performances align with the director's vision. 
We utilize SiA to compile a diverse dataset featuring Sophia in a variety of performances, including recreations of iconic scenes from classic films for selected clips complemented by the corresponding environmental lighting. Additionally, certain segments are captured from multiple angles, resulting in a collection of 3D Gaussian Splatting (3DGS) sequences.
%
Recall that Sophia only presents a single identity, we therefore apply re-enactment techniques to further enrich the appearances and styles of the performance while preserving realism (Fig.~\ref{fig:dataset} and the supplementary video). The complete SiA dataset, the first of its kind, will be disseminated to the research community. Our comprehensive user studies on Sophia's dynamic performances have unveiled that a majority of viewers found these performances both acceptable and enjoyable, generally reducing the uncanny valley effect. This indicates a positive audience response towards the use of humanoid robotics in virtual acting. Furthermore, the immersive lighting in dynamic clips received high praise for its similarity to natural lighting and its ability to mirror professional film standards, thus significantly enhancing the aesthetic appeal of Sophia's performances. At the same time, they unanimously recognize it is a robot performing and do not confuse it with a human performance. 

\begin{figure*}[ht]
  \centering
  \includegraphics[width=\linewidth]{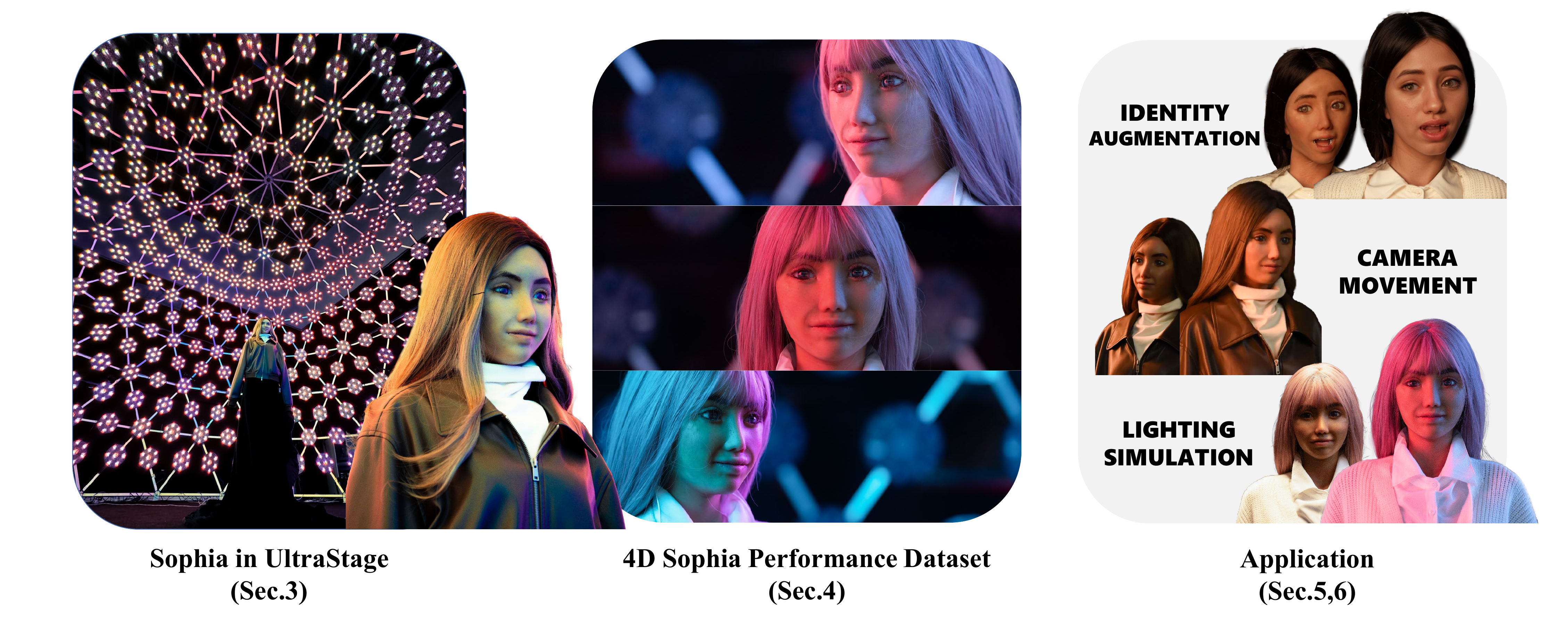}
  \caption{The overview of Sophia-in-Audition(SiA). We deploy the humanoid robot Sophia as a virtual performer inside the UltraStage (Sec.~\ref{sec:sophia-in-audition}) composed of a controllable lighting dome coupled with multi-camera video shooting. We use SiA to collect a comprehensive dataset with Sophia in immersive environment lighting (Sec.~\ref{sec:dataset}). We call this setup Sophia-in-Audition or SiA which allows for simultaneous controls over performance, lighting, and camera movements (Sec.~\ref{sec:identity_augmentation},~\ref{sec:virtual_lighting_cam_move}).}
  \label{fig:overview}
\end{figure*}

Overall, our contributions include:
\begin{itemize}
\item We experiment with the possibly first virtual production setup with a robot performer, Sophia-in-Audition, that has the Sophia robot perform within the UltraStage environment under the multi-view capture setup. 
\item We develop a complete set of tools for controlling Sophia's facial motion and expression while preserving emotion, emulating environment lighting in classic clips, conducting camera moves via multi-view rendering, and enriching the appearance using generation techniques. 
\item We gather and disseminate the first multi-view robot performance video datasets with dynamic lighting. We also conduct a comprehensive user study to evaluate the performance and the immersive lighting of SiA. Overall, SiA serves as an excellent virtual audition solution, to emulate trial shots that potentially allow filmmakers to coordinate various elements (lighting, camera movement, and performance) before committing to the full shoot.
\end{itemize}

%% file: sec/2_related.tex
\section{Related Work}
\label{sec:related}
As the first attempt to deploy a robot performer within a controllable lighting environment under the multi-view capture setting, SiA employs the latest advances in virtual production to harness the power of imaging, modeling, rendering, and generation. 

\paragraph{Virtual Environments.} A key component in virtual production is virtual environments to allow direct blending between the physical and digital worlds. For long, chroma keying has been the dominating method in virtual production that involves shooting scenes against a green or blue backdrop, which is later replaced with digitally generated content that would be otherwise challenging or expensive to film. A major drawback is the mismatch between the lighting of the actual shoot and the virtual environment. Traditionally, this was addressed by establishing specialized lighting teams or applying post-production relighting \cite{wenger2005performance, chabert2006relighting, peers2007post}. Significant advances have been made in post-capture relighting, ranging from physical-based rendering \cite{pharr2023physically} to data-driven relighting \cite{bi2021deep, zhang2021neural, pandey2021total} and to generation-based solutions \cite{rombach2022high}. The best effects, however, are still on static images. Handling video sequences using post-production relighting is more challenging as existing techniques cannot sufficiently maintain temporal consistency and hence commonly cause flickering.

Virtual Production Studios (VPS) resort to LED walls \cite{VPS} that are composed as simple as single walls and as complex as curved structures enveloping entire scenes. Real-time rendering is then employed to project rendered contents onto the LED walls along with camera movements. This allows directors, actors, and cinematographers to view and interact with the scene as well as make scene modifications and transitions easily manageable. Yet, VPS have their own limitation, e.g., the proximity of LED walls to actors limits their ability to simulate distant light sources. More importantly, they are prohibitively costly when constructed in large sizes. A more affordable alternative is the Light Stage \cite{debevec2000acquiring, hawkins2001photometric, debevec2002lighting, wenger2005performance, debevec2012light}, pioneered by Debevec and his team. A Light Stage is a dome environment densely populated with colored light beads, allowing the simulation of light sources from any direction in the environment. Latest extensions such as the UltraStage \cite{zhou2023relightable} employ light beads with high-frequency controls to enable more realistic contrasts of light and dark, creating both hard and soft lighting effects on actors' faces. Their light sources can be freely positioned in 360 degrees, enabling a balanced distribution of primary and secondary light sources where the color of each bead can be precisely adjusted.


\paragraph{Virtual Performers.} While existing virtual production has relied on real performers, there is an increasing interest in adopting virtual performers, from entirely computer-generated characters to digital doubles \cite{bryndin2019human, dixon2005digital} or de-aged versions of real actors \cite{benjamin_button, the_irishman} in live-action movies. In this paper, we explore the third type, a robot. A major driving force behind the virtual performer is motion capture (MoCap), to enable transferring body movement and facial expressions from real performers to virtual ones. For the former, State-of-the-art MoCap solutions \cite{van2018accuracy, vicon} use infrared markers and a stationary camera system where actors donned in suits covered with reflective markers perform while cameras track the reflected infrared light. Recent learning-based techniques have given rise to markerless MoCap \cite{nakano2020evaluation, rempe2021humor, berger2011markerless, knippenberg2017markerless} to achieve comparable quality and we refer the readers to the comprehensive survey \cite{desmarais2021review} for more details. In a similar vein, facial MoCap \cite{vicon_facial, lin2005mirror, darujati2013facial, bickel2007multi} allows an actor to wear a helmet equipped with a camera to record their facial movements. This setup provides animators with highly detailed data, enabling them to create lifelike facial expressions for virtual characters. To avoid the use of bulky and cumbersome helmets, Apple's ARKit \cite{Apple_arkit} provides unprecedented capabilities for tracking human motion and facial expressions without the need for physical markers. In this paper, we demonstrate how to extend these approaches to the Sophia robot.  

Virtual performers are also stimulated by the latest advances in learning-based identity re-enactment. GAN-based techniques in the early days such as Deepfakes \cite{deepfakes} and Deepfacelab \cite{perov2021deepfacelab} enable the likeness of one person in an existing image or video is digitally superimposed with the face of another person. More recently, diffusion-based methods such as ControlNet \cite{controlnet}, IP-Adapter \cite{ye2023ip-adapter}, and DreamBooth \cite{ruiz2022dreambooth} showcase the potential of leveraging the great power of diffusion-based pretrained image and video generation models, such as stable diffusion \cite{rombach2022high} and AnimateDiff \cite{guo2023animatediff}, to produce high-quality images and videos with control of desired person identity. Tools like ReActor \cite{sd-webui-reactor} further bridge the gap between traditional deepfake methods and these novel diffusion-based approaches, signifying a continuous community effort in the creation of realistic virtual performers. The biggest challenge though is the trade-off between realism and temporal consistency. Traditional deepfake methods offer good temporal consistency but lack realistic and natural portrayal. Conversely, stable diffusion-based methods provide superior static quality but struggle with maintaining temporal consistency. 

Deploying humanoid robots to serve as virtual performers is still in its cradle and it requires seamless integration of robot control, animation, and rendering. We brace ourselves to conduct the first such trials. We observe that existing social robots have focused on convincingly emoting and expressing a wide range of human-like emotions and movements (facial expressions, body language, natural-sounding speech and voice synthesis, etc). They hence serve as perfect candidates for virtual performers in our setup. Exemplary models include Geminoid DK \cite{GeminoidDK}, Erica \cite{Erica}, HRP-4C \cite{HRP-4C}, and Sophia \cite{Sophia}. Among them, Sophia stands out among these as one of the most expressive and realistic. Yet Sophia's control differs from existing CG models where deploying facial MoCap and conducting identity re-enactment require elaborate designs. We address these key challenges, jointly with virtual relighting and camera moves, to preserve the expressiveness and realism of static and dynamic expressions. 

%% file: sec/3_system.tex
\begin{figure}[ht]
  \centering
  \includegraphics[width=\linewidth]{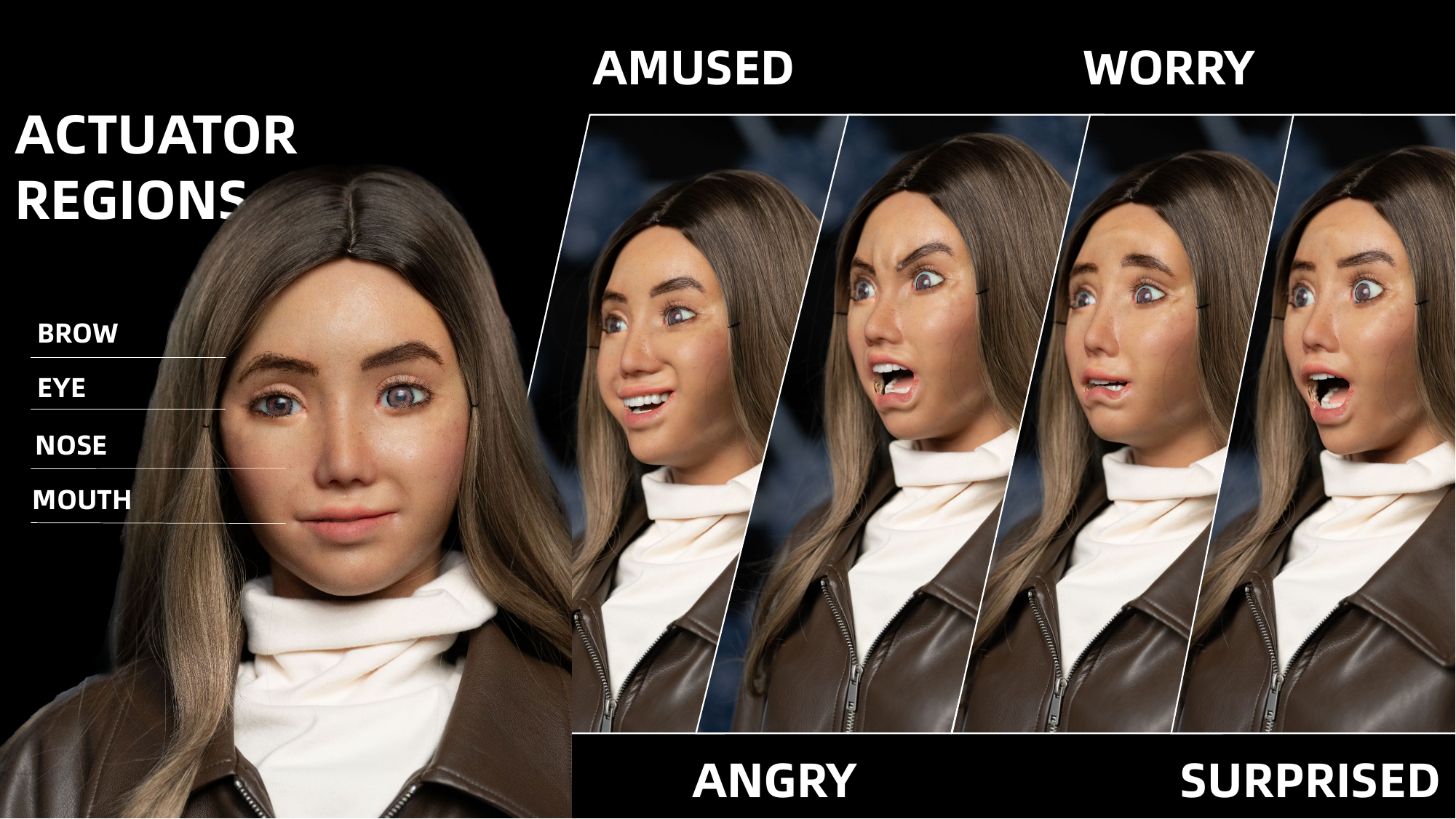}
  \caption{Virtual Performer Sophia and Her Facial Expression Samples. The brow region features 5 motors allowing for eyebrow lifting, while the eye and eyelid area is equipped with 11 actuators for blinking and eye movement. Additionally, the nose region contains 2 motors, and the mouth area includes 14 motors that control movements of the mouth, tongue, jaw, and surrounding muscles, enabling a wide range of expressive capabilities.}
  \label{fig:sophia_motor}
\end{figure}

\section{Sophia-in-Audition}
\label{sec:sophia-in-audition}
Our Sophia-in-Audition (SiA) system (Fig.~\ref{fig:overview}) exploits the latest advances in humanoid robots and virtual production. In this section, we address specific challenges to incorporate the Sophia robot as a virtual performer. We also discuss how SiA can serve as a virtual audition solution by enabling simultaneous controls over the performance, lighting, and camera movement.

\subsection{Sophia the Robot}
\label{subsec:sophia_the_robot}
Sophia \cite{Sophia}, an advanced robot unveiled by Hanson Robotics since 2016, stands as an icon of humanoid robotics. The most notable feature of Sophia is her capability to display highly complex as well as nuanced facial expressions, marking a significant advancement in human robot interactions. Sophia embodies the potential of a sophisticated virtual performer also because of her integrated platform including advanced perception, interaction, actuation, and control systems. These systems, coupled with the latest Large Language Models \cite{openai2023chatgpt} and image generation techniques \cite{rombach2022high}, enable Sophia to perform as a virtual actor.

\paragraph{Motion Controls.} To briefly reiterate, Sophia employs a sophisticated motion system to emulate human facial expressions. This includes motor-controlled cords anchored underneath her skin, which is made of a flexible, human-like rubber material called ``Frubber" \cite{oh2006design}. The core of Sophia's facial expressiveness lies in the precise coordination of actuators within her head and face. These actuators, responsible for a total of 33 Degrees-of-Freedom (DoF), support the emulation of a broad spectrum of human-like expressions. They function similarly to human facial muscles, allowing Sophia to demonstrate nuanced emotions such as joy, grief, curiosity, confusion, contemplation, sorrow, and frustration, as shown in Figure~\ref{fig:sophia_motor}. For example, the 5 motors in her brow and forehead area enable Sophia to lift her eyebrows, a movement reminiscent of the human frontalis muscle, crucial for expressing surprise or curiosity. In the region of her eyes and eyelids, there are 11 actuators that facilitate eye blinking and eye movement, mirroring the functions of the human orbicularis oculi muscle and medial rectus muscle that play a vital role in non-verbal communication. An additional group of 14 motors represent the mouth, tongue, and the surroundings of her mouth and lips, allowing for intricate mouth movements and other nuanced movements such as smiling and frowning, analogous to the zygomaticus major and depressor anguli oris muscles in humans. \cite{Uldis2017}

It is critical to note that, despite these advanced facial motor systems, Sophia cannot yet fully replicate the range, subtlety, and fluidity of facial movements of a real performer. For example, the mechanical actuators, while versatile, cannot fully emulate the subtlety, continuous and smooth motion characteristic of natural human expressions. This difference is particularly notable when comparing her expressions to the wide range of human facial movements largely used in the computer graphics community, known as BlendShapes \cite{blendshape}, which are more nuanced and varied. Additionally, integrating BlendShape control methods into Sophia's physical system presents challenges, as these digital animation techniques do not translate directly to the mechanical movement of robotics. This gap highlights the complexities of achieving lifelike expressiveness in humanoid robots and underscores the ongoing challenges in the field. In subsequent sections, we will discuss potential solutions to address and narrow this gap, moving closer to more lifelike and fluid robotic expressions.

\paragraph{Facial Expression Emulation.}

Hanson Robotics provides a comprehensive SDK in ROS to offer three distinct control methods. The first is the direct actuator control API, which allows precise, individual actuator manipulation, providing the ability to achieve specific poses at determined times. This method is crucial for detailed and customized movements, particularly for expressions and gestures that require fine-tuned control. The animation API facilitates more fluid, BlendShape-like movements, involving coordinated actuation for part or all of Sophia's face. This approach is integral to creating lifelike facial expressions, drawing from a rich library of expressions, animations, and visemes. The prioritization within this system ensures that essential expressions, like visemes for speech clarity, take precedence over other animations. Finally, the performance API offers a higher-level control, orchestrating complex performances that combine various pre-created animations and expressions \cite{SophiaDocs}.

There are two primary approaches to control Sophia's facial motion. The first is a keyframe-based animation, which, while precise, demands an extensive amount of labor. Specifically, it involves meticulously creating frame-by-frame animations to dictate Sophia's movements, requiring significant time and effort to achieve realistic and fluid motion. The second approach utilizes an off-the-shelf method to drive Sophia's movements directly from a human performer. It is less labor-intensive and allows for more dynamic and natural performance, as it captures the spontaneity and subtlety of human motion. In our implementation, we extend the second approach by integrating Apple's ARKit \cite{Apple_arkit} to first extract motion from human performers and then use the results to guide more vivid and lifelike performances. Specifically, we create and optimize the mapping matrix between the ARKit BlendShapes parameters and the actuator parameters of Sophia, so that driving Sophia using ARKit's capture signal of Sophia's movement will always produce the same movements. This integration allows us to harness lifelike motions from Sophia, facilitated by the advanced capabilities of ARKit.

Our solution enables Sophia to replicate a substantial portion of the facial movements captured by the ARKit BlendShapes, despite her actuators covering about only half of these BlendShapes. There are a number of inherent limitations of our current approach. First and foremost, the BlendShape movements captured by ARKit still differ from the movements produced by Sophia's motor-driven virtual muscles. Additionally, the motors' physical power is rather small which subsequently constrains the rapidity and range of facial expression changes. In our experiments, we find these limitations do weaken Sophia's ability to fully replicate certain facial expressions captured by ARKit (e.g., big smiles or minute expressions). Yet, overall, it suffices the need as a preliminary study on robot performers: in the end, as aforementioned, we do not intend to replace real actors; rather, Sophia serves as an excellent "audition" performer, to allow changes to the dialogue, the lighting, and the camera moves. 

In the process of animating Sophia using ARKit, we also experimented with two primary sources of motion capture: performances by human actors and recorded videos or film clips of famous actors. We observe that the former, involving skilled actors, tends to yield more dynamic and authentic results, capturing the subtleties of live expression. However, it requires experienced actors (in our case, we recruited top students from theater academies) to replicate the vividness of the performance. Using pre-recorded videos or films (e.g., legacy clips) is more convenient but faces several challenges. For example, many iconic clips use tailored lighting where motion extraction can be particularly challenging; side views are also much more difficult to handle than front views. We conduct de-lighting by from the estimated environment light (Sec.\ref{subsec:3-2_virtual_lighting}) to address the first challenge and use stable diffusion based image completion to partially mitigate the second. 

\begin{figure}[ht]
  \centering
  \includegraphics[width=\linewidth]{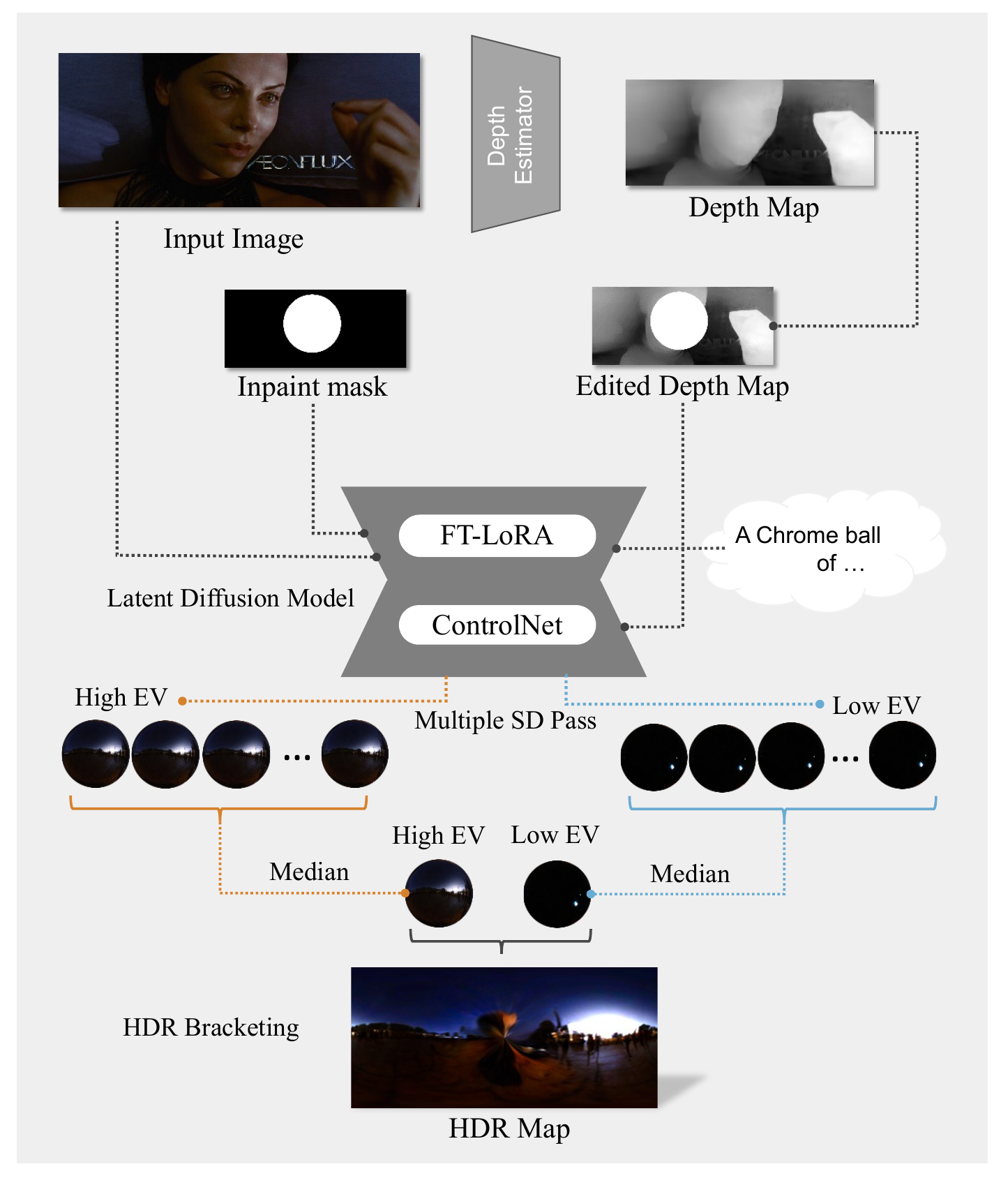}
  \caption{
  Lighting Estimation Pipeline. The estimation begins with an input image undergoing depth estimation and subsequent depth map editing, which, alongside an inpainting mask, feeds into a latent diffusion model with fine-tuned LoRA (FT-LoRA) to generate chrome ball reflections at various exposure values. These are then median combined using HDR bracketing to synthesize a comprehensive HDR map. 
  \label{fig:lighting_estimation_pipeline}
  }
\end{figure}

\begin{figure*}[ht]
  \centering
  \includegraphics[width=\linewidth]{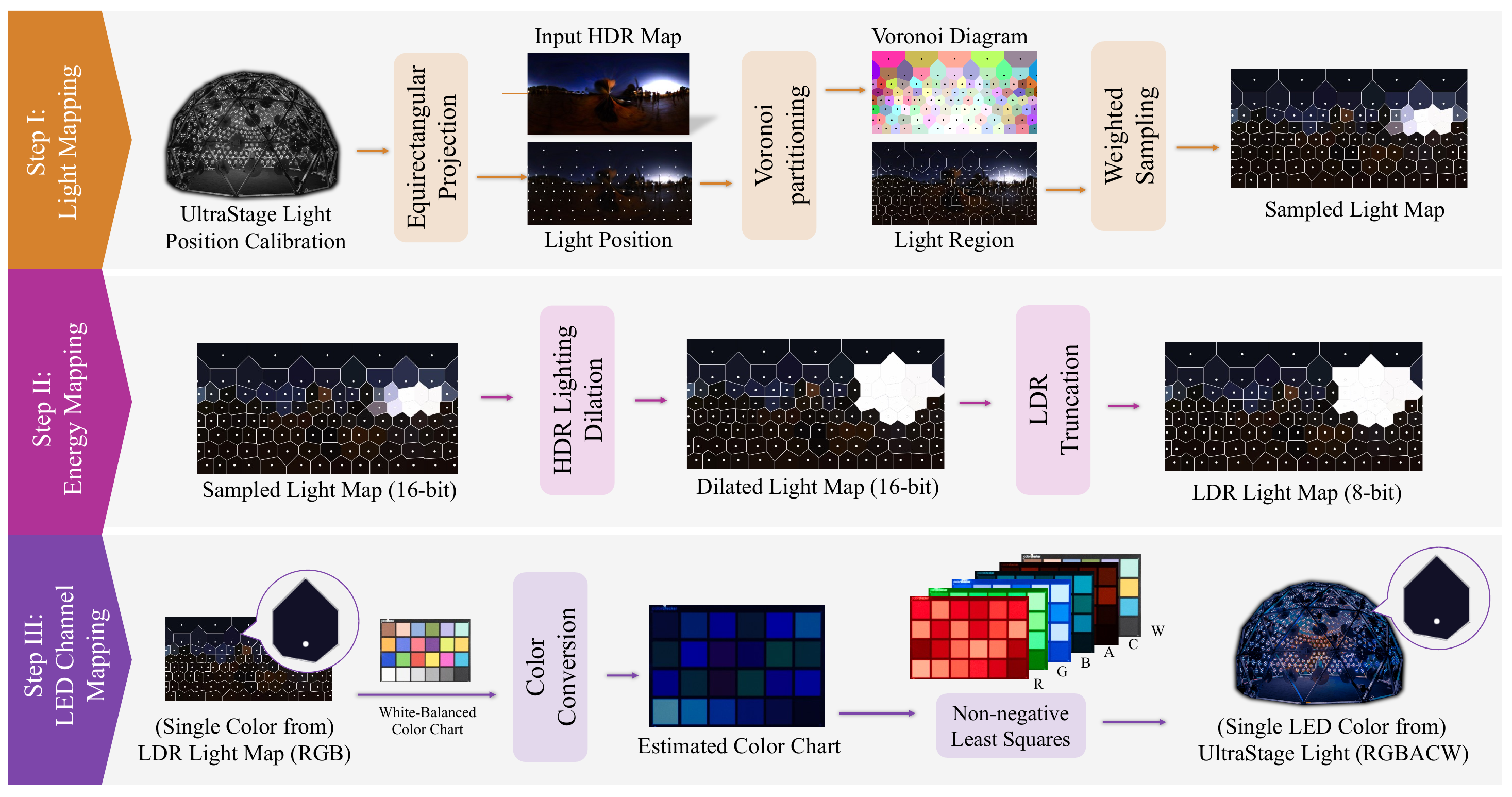}
  \caption{
    Lighting Reproduction Pipeline. For lighting reproduction, the pipeline entails a three-stage process within the UltraStage: first, calibrating and projecting light positions onto the HDR map, followed by Voronoi partitioning and weighted sampling to capture light regions and colors. Next, the sampled HDR light map is down-converted to LDR while preserving energy through dilation of overexposed areas. The final stage involves mapping the three-channel light map onto a six-spectrum LED arrangement using a color chart and non-negative least squares method, producing a precise LED light map that is utilized within the UltraStage for accurate environmental lighting emulation.
  \label{fig:lighting_reproduction_pipeline}
  }
\end{figure*}

\begin{figure*}[ht]
  \centering
  \includegraphics[width=\linewidth]{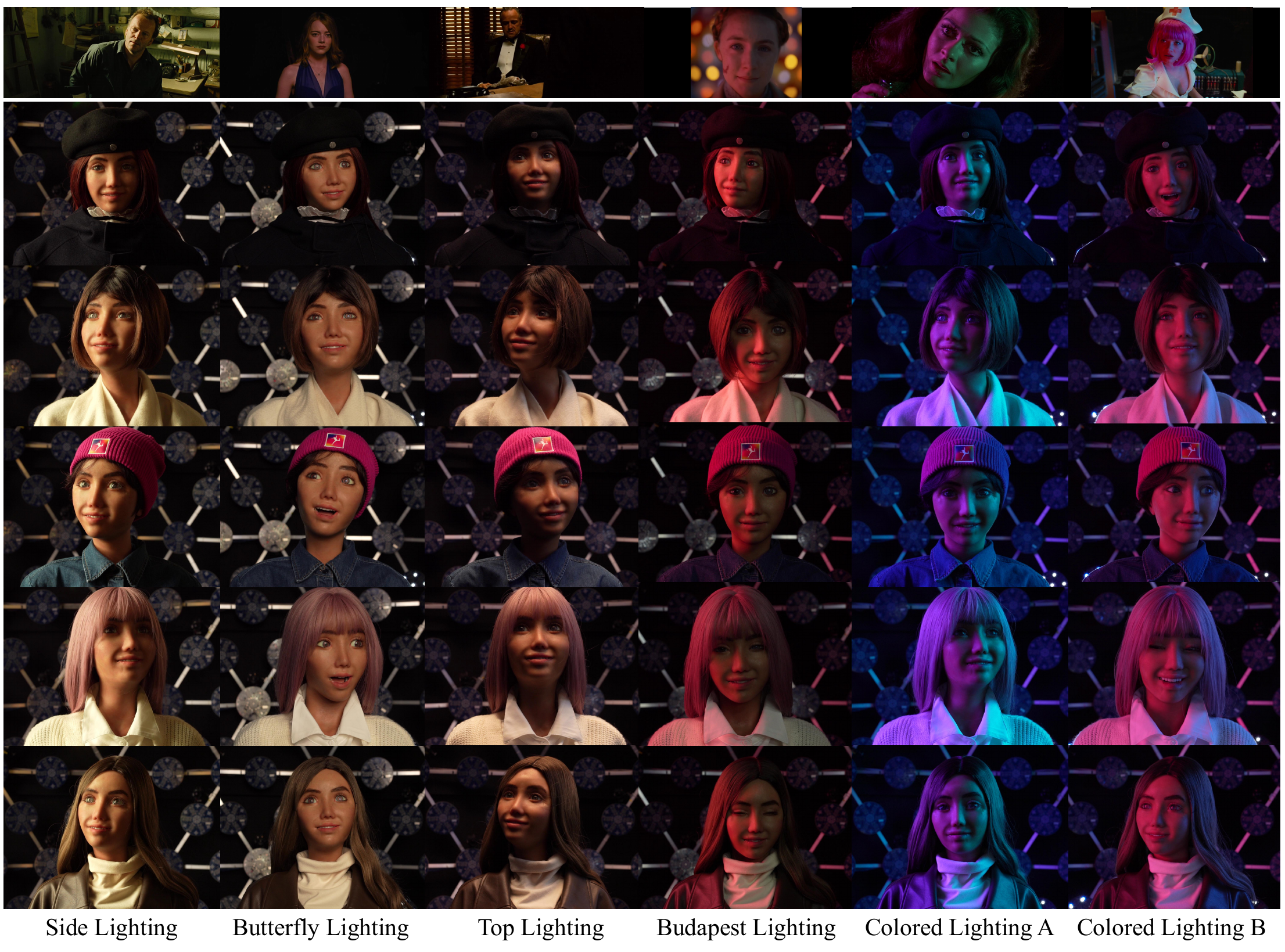}
  \caption{The SiA Dataset. Introducing an innovative virtual production dataset featuring performer Sophia, this collection encompasses a wide array of elements, including diverse cinematic lighting setups, an assortment of costumes, and a spectrum of expressions. Highlighted cinematic lighting examples include side, butterfly, and top lighting, alongside renowned setups derived from classic film scenes. The first row displays corresponding film scenes that serve as the basis for our lighting estimation (Images courtesy of \textit{Across the Universe} \cite{AcrosstheUniverse}, \textit{La La Land} \cite{LaLaLand}, \textit{The Godfather} \cite{TheGodfather}, \textit{The Grand Budapest Hotel} \cite{budapest}, \textit{Beyond the Valley of the Dolls} \cite{BeyondtheValleyoftheDolls}, and \textit{The Zero Theorem} \cite{TheZeroTheorem}).}
  \label{fig:dataset}
\end{figure*}

It is particularly worth mentioning that our recovered facial expressions from live performances or videos do not sufficiently reflect the actual emotion of the performance. The problem is further magnified due to the limited motor motion range of Sophia. As a result, the facial expression performed by Sophia can fail to convey the actual emotion. To address this challenge, we adopt a preset of strong emotional expressions such as amused, angry, engaged, fearful, and sad. Instead of directly using the recovered facial motion (e.g., from ARKit) to guide Sophia, we first classify the emotion state for each keyframe using GPT4-V from ChatGPT \cite{openai2023chatgpt} and then blend the corresponding preset expression with the recovered ones to drive Sophia. Post-editing of the captured signals was also occasionally manually fixed to enhance the emotion as well as to fit the motor motion range.

Another challenge in translating ARKit signals to motor parameters is the occurrence of stuttering. This issue may arise from frame loss during ARKit capture or from motor limitations when handling rapid changes in facial expressions. To mitigate the artifacts, we adopt a two-step approach. Firstly, we conduct interpolation and key-frame editing of ARKit capture signals to produce smoother motion transitions for the motors. If it still fails to sufficiently smooth the motion, we employ a frame-by-frame capture of the Sophia performance. This method focuses on capturing the steady state of the motors, making it less affected by transient signals from the actuators and ensuring smoother movement. Supplementary videos show that such approaches greatly improve the smoothness and fluidity of Sophia's performance. 

\subsection{UltraStage: Virtual Lighting}
\label{subsec:3-2_virtual_lighting}
\input{sec/3-2_virtual_lighting}

%% file: sec/3-2_virtual_lighting.tex
As we aim to have Sophia perform as if in an audition, it is essential that we also integrate two additional components in virtual productions: virtual lighting and virtual camera movement. Instead of using an LED wall, we resort to the Light Stage type solution called the UltraStage \cite{zhou2023relightable}. The UltraStage is a 10-meter diameter spherical dome, simulating approximately three-quarters of a sphere. Its ground-level diameter spans 8 meters, providing ample space for performance and filming. The dome is equipped with 480 evenly distributed light panels, enabling the simulation of light sources from any direction in a 360-degree range. Each panel consists of 46 light beads, featuring six spectral types: Red, Green, Blue, Amber, Cyan, and white. This six-spectrum LED system offers a broader color gamut than traditional RGB Light Stage systems, achieving more realistic environmental light colors. Each bead's brightness is controllable with an 8-bit input range. The white lights can compensate for the brightness dynamic range of 8-bit, enhancing overall lighting contrast to approximate HDR environmental lighting. Additionally, each light bead is programmable for individual control. All panels are managed via a network DMX512 signal, allowing for real-time dynamic lighting at a high frame rate of 120Hz.

\begin{figure*}[ht]
  \centering
  \includegraphics[width=\linewidth]{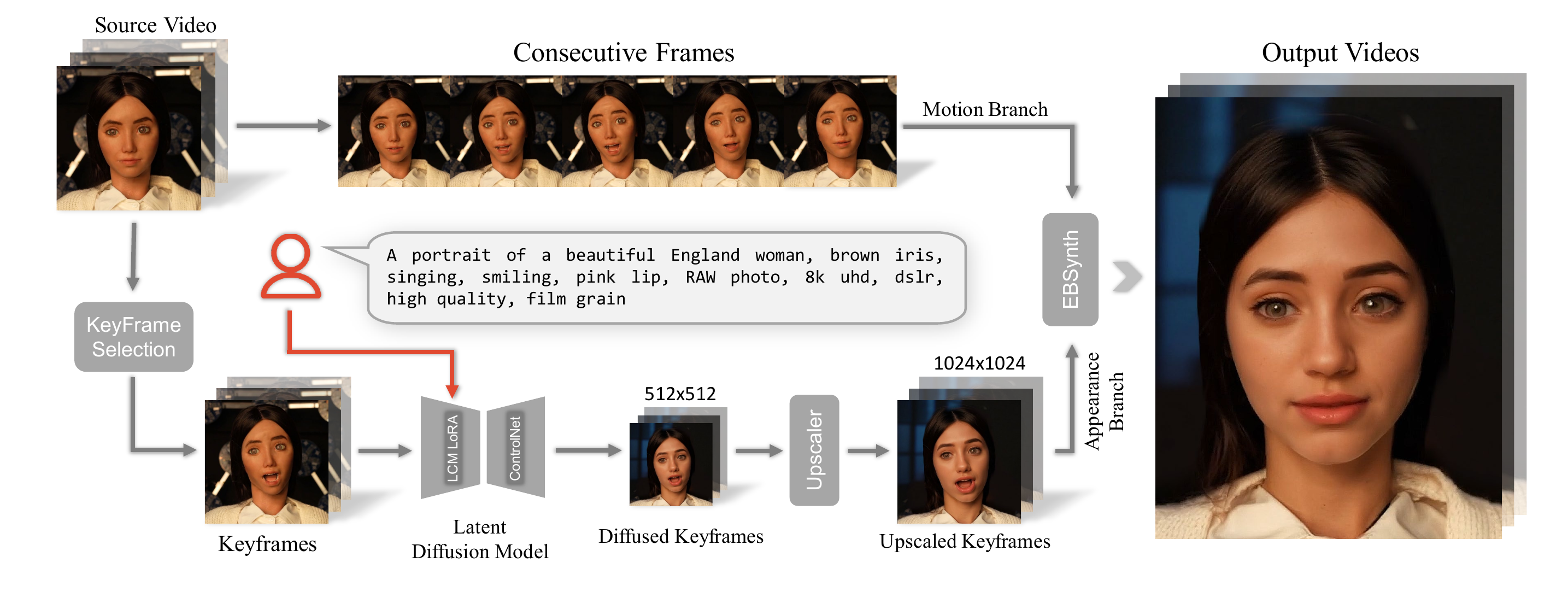}
  \caption{Identity Augmentation Pipeline. We designed a dual-track video generation method that separates the appearance and motion aspects of the process. We start with selecting key frames from Sophia's performances, ensuring these frames effectively represent the video's local segments. This step is crucial in capturing the essence of her performance. We then apply SD to these keyframes, guided by textual prompts aimed at enhancing and enriching Sophia's identity. This approach allows us to significantly improve her appearance while preserving the original video's major motion, temporal consistency, facial expressions, and environmental lighting. In comparison, direct generation of the entire video clip that maintains high quality across these aspects is extremely challenging whereas our keyframe-based solution is more tractable and very easy to use.}
  \label{fig:identity_aug_pipeline}
\end{figure*}

\paragraph{Lighting Estimation.}
An exemplary task in SiA is to have Sophia replicate an iconic clip. Recall that the lighting of such clips encompasses the cinematographer's intricate design and artistic vision. It enhances actors' appearances, sets the emotional tone, and advances the narrative. Cinematographers meticulously consider aspects like the direction, color, temperature, and quality (diffuse or hard light) of each light source, as well as the balance between key and fill lights. Therefore, it is crucial to first reproduce these exquisitely designed movie lighting used in corresponding film segments. 

Estimating environmental light from a single image remains a highly challenging task for computer vision. As illustrated in Fig.~\ref{fig:lighting_estimation_pipeline} We followed DiffusionLight \cite{Phongthawee2023DiffusionLight}, adopting a diffusion-based scheme \cite{rombach2022high} integrating depth estimation, inpainting, and exposure modeling. Starting with a movie scene screenshot, we first estimate its depth map using learning-based monocular depth estimation \cite{ranftl2020towards}. Next, hypothesize that there is a probe (i.e., a mirror sphere) at the center of the image. We carve out this region, marked by a white circle, and then apply using stable diffusion and iterative techniques guided by prompts to refine it. ControlNet \cite{controlnet} is used in the inpainting step to generate a chrome ball in the masked area. The initial light probe is an LDR image, which we enhance to HDR quality by training a LoRA \cite{hu2021lora} model on simulated renderings of metal spheres at different exposures. Multiple LDR light probes are merged to create an HDR light probe using a staged exposure synthesis method \cite{tufford2012bracketing, gearing2004bracketing}. This HDR environmental light is then unwrapped into an equidistant cylindrical expansion for practical application.

\paragraph{Lighting Reproduction.}
Once we estimate the HDR environment map, we set out to emulate the corresponding lighting conditions with the UltraStage system, as shown in Fig.~\ref{fig:lighting_reproduction_pipeline}. This is achieved by first mapping the HDR environmental light onto a spherical surface. Using the 3D positions of 480 lights, the sphere is divided into 3D Voronoi \cite{aurenhammer1991voronoi} segments, with each light panel having a unique Voronoi cell. Each pixel of the HDR image is assigned to a cell. Next, we calculate the irradiance of each light panel using cos-weighted sampling, ensuring the total energy remains consistent. The dynamic range is managed by using Gaussian blur to smooth extremely bright areas, followed by HDR dilation to distribute excess irradiance among surrounding panels \cite{debevec2022hdr}. Finally, to replicate the real-world physics of light, we convert the RGB values of the HDR map into coefficients in six different LED spectra, derived using a 24-color card and a non-negative least squares optimization method \cite{legendre2016practical}. This process ensures that the recreated lighting accurately represents the original HDR environmental lighting. We have further designed an interactive lighting control interface, allowing for the specification of individual LED parameters. This facilitates the fine-tuning of each environmental light source's direction, brightness, and color, ideal for creating imaginative light sources and giving directors and artists the freedom to achieve their desired outcomes.

\begin{figure*}[ht]
  \centering
  \includegraphics[width=\linewidth]{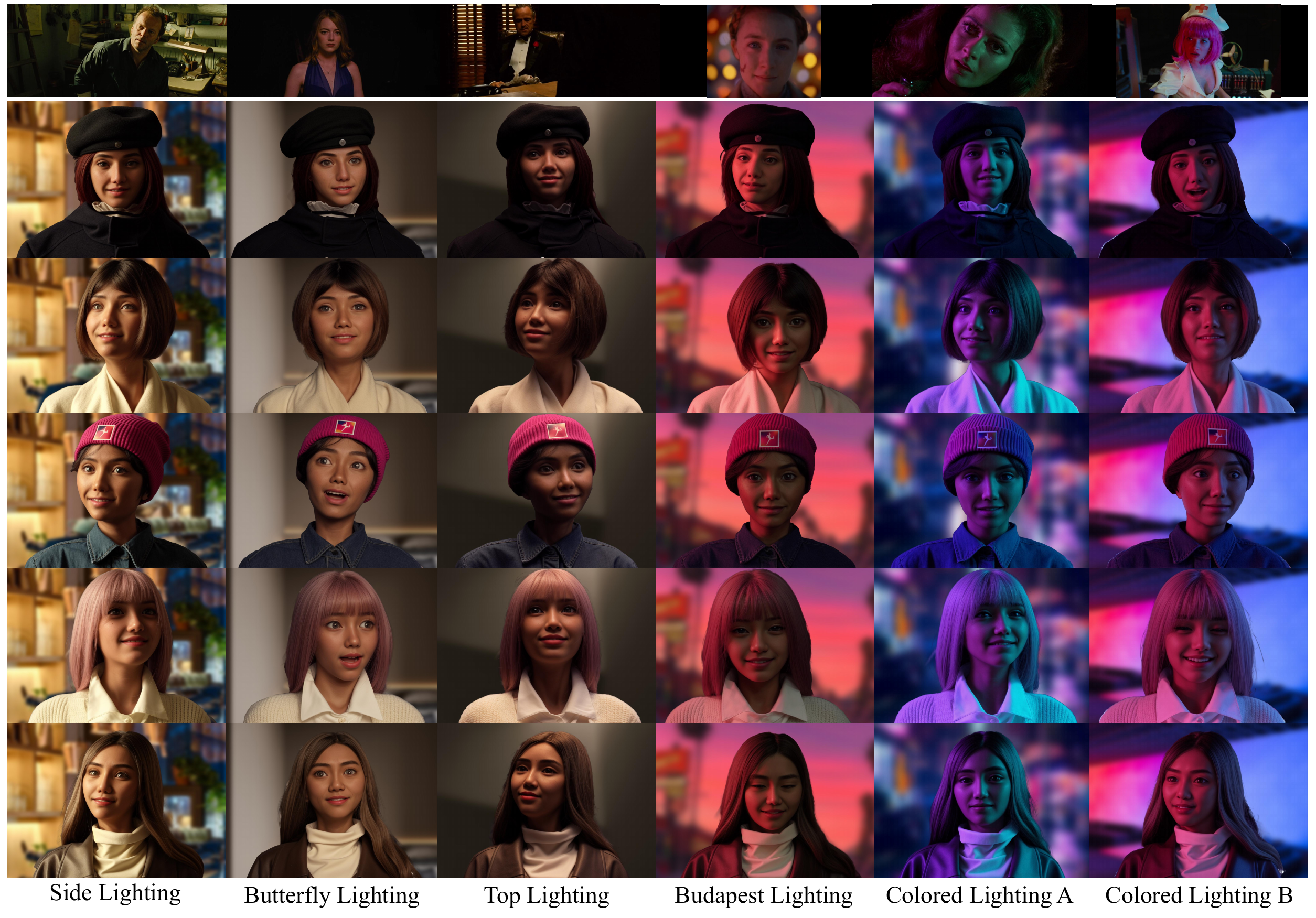}
  \caption{Immersive Lighting Results. This collection showcases characters in five unique costumes, each bathed in one of six cinematic lighting arrangements (the first row). Thanks to the application of Stable Diffusion, we attain more realistic depictions of ``Frubber" skin, complete with authentic highlights and reflections. The wigs are finely tuned to reflect the subtle intricacies of the surrounding lighting environment. Notably, UltraStage's emulation of top and side lighting preserves the depth of shadows and stark contrasts in luminance with remarkable accuracy. The hues achieved through our lighting replication faithfully mirror those of traditional cinema, effectively capturing the emotional resonance, atmosphere, and distinctive aesthetic of classic film scenes (Images courtesy of \textit{Across the Universe} \cite{AcrosstheUniverse}, \textit{La La Land} \cite{LaLaLand}, \textit{The Godfather} \cite{TheGodfather}, \textit{The Grand Budapest Hotel} \cite{budapest}, \textit{Beyond the Valley of the Dolls} \cite{BeyondtheValleyoftheDolls}, and \textit{The Zero Theorem} \cite{TheZeroTheorem}).}
  \label{fig:light_similation_result}
\end{figure*}

Compared with image-based relighting \cite{sun2020light, bi2021deep, zhang2021neural, pandey2021total, pharr2023physically}, physically simulated lighting using UltraStage has several advantages. Image-based relighting, being fundamentally 2D, struggles with temporal and spatial consistency due to its inability to fully incorporate 3D information. This limitation makes it challenging to replicate dynamic changes in lighting over space, an essential aspect of realistic film lighting. UltraStage, using a 3D dome environment, successfully overcomes these challenges by accurately simulating correct 3D lighting. Another critical limitation of image-based relighting is its difficulty in decoupling real material colors from lighting colors. This process requires solving an inverse rendering problem to separate object normal and albedo, among other material properties. Additionally, image-based methods struggle with complex materials, only effectively handling simple diffuse or specular BRDFs. They fall short in situations like subsurface scattering in skin, complex reflections in hair, multiple scattering in furry materials, or handling objects with complex geometry and self-occlusion. UltraStage, as a comprehensive environmental light simulator, eliminates these issues and conveniently reproduces dynamic and realistic lighting, particularly the expressive lighting seen in films. Fig.~\ref{fig:light_similation_result} and the video show several examples. 

\paragraph{Multi-camera Setup.}
Previous Light Stage systems generally used a sparse (1 to 8) set of cameras, as the focuses were conducting relighting or photometric stereo. UltraStage, in contrast, is equipped with a multi-camera capture system that conducts synchronized shooting with 32 cameras, with each camera capturing intricate details of performances from different angles. The multi-view nature of the capture further supports free-viewpoint cinematography, either with neural rendering \cite{mildenhall2021nerf} or the latest 3D Gaussian Splatting \cite{kerbl20233d}. Details on using multi-view rendering for camera movement can be found in Section~\ref{sec:virtual_lighting_cam_move}. 

\begin{figure*}[ht]
  \centering
  \includegraphics[width=\linewidth]{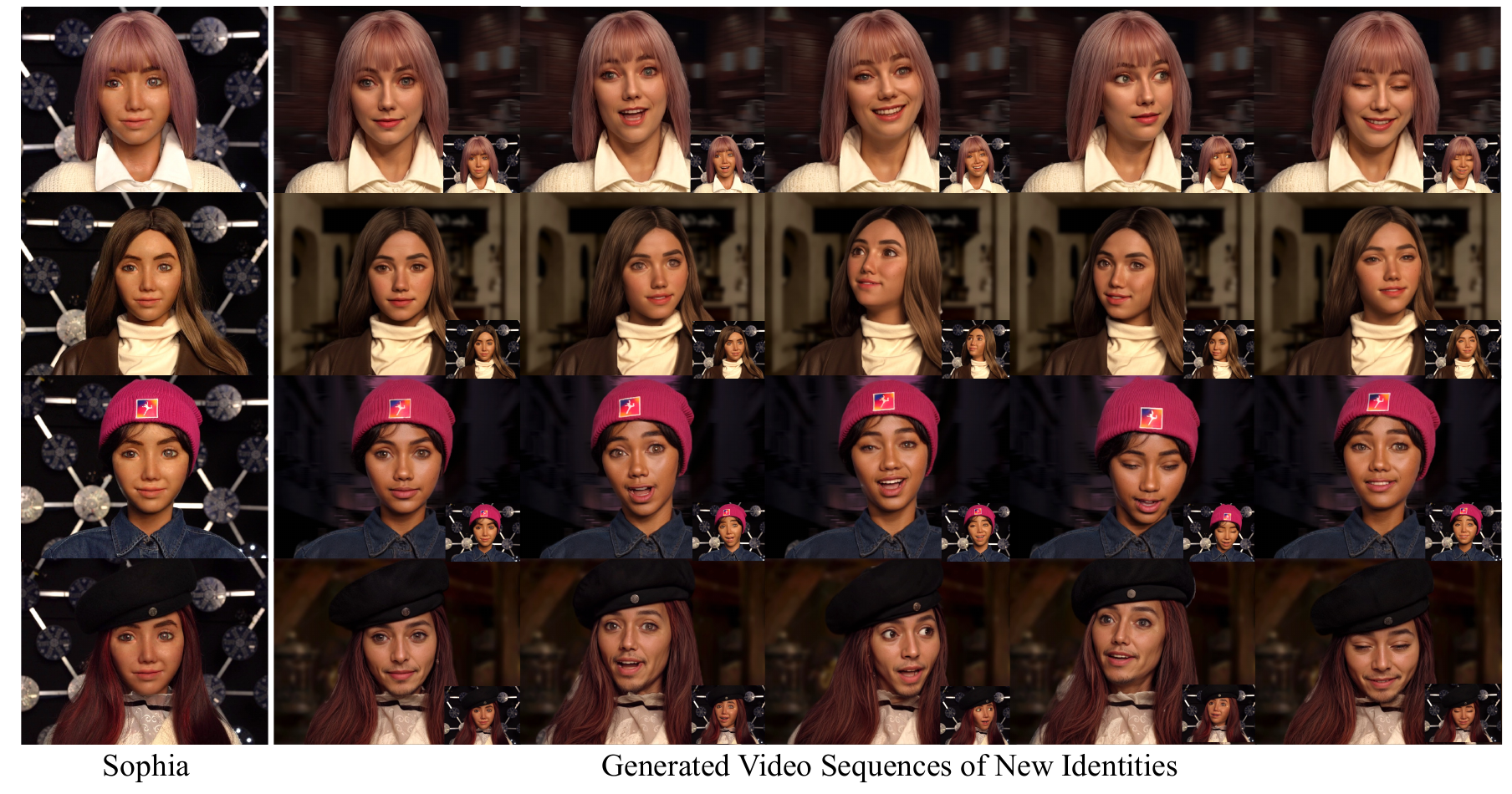}
  \caption{Identity Augmentation Results. Beginning with authentic video captures of Sophia (an example frame in the first column), our methodology produces temporally consistent outcomes (subsequent columns) for a specified generated identity, illustrating our system's proficiency in augmenting identity diversity. Each row employs a unique prompt to dictate the identity's appearance, ensuring identity consistency across frames while each row introduces a distinct identity. Notably, the resultant sequences exhibit enhanced realism, particularly in skin reflectance, shadows near the nose, and eye highlights, attributable to the advanced generative capabilities of Stable Diffusion. In each generated image, the input image of Sophia is in the bottom right corner for reference.}
  \label{fig:identity_aug_result}
\end{figure*}

%% file: sec/4_dataset.tex
\section{The SiA Dataset for Virtual Production}
\label{sec:dataset}
Before we showcase the applications of SiA, we first introduce a novel virtual production dataset to be shared with the community. This dataset encompasses 50 unique video segments that capture Sophia's performances in UltraStage.
The collection features a mix of single-view captured clips and, for some segments, multi-view captures that include up to 32 views to illustrate the feasibility of virtual camera movement. Accompanying these segments are corresponding environment maps (including the ones estimated from well-known clips), as well as facial motion capture (mocap) data employed to drive Sophia.


\paragraph{Dataset Content.} 
The SiA dataset is a rich compilation of diverse elements.
It encompasses a variety of lighting conditions that have been broadly adopted by directors in a plethora of film segments, including Rembrandt lighting, butterfly lighting, side lighting, top lighting, colored split lighting, dynamic colored lighting, and a challenging lightning case. Figure \ref{fig:dataset} showcases some representative lighting styles.

In the first column of Fig.~\ref{fig:dataset}, we illustrate the dramatic effect of side lighting, notable for its stark contrast and the distinctive delineation it creates across the face, thereby accentuating facial contours and depth. Realizing such a nuanced lighting effect necessitated meticulous adjustment of the light source's direction alongside the precise positioning of the actor's face.
We employed a combination of estimated HDR environmental lighting mapped onto the UltraStage and an interactive, real-time lighting control UI. This setup allowed us to fine-tune the light's color and direction via network signals, ensuring that the lighting on the actor's face was as close to perfect as possible. The diverse class of lighting enriches the dataset with cinematic quality and versatility.


\begin{figure*}[ht]
  \centering
  \includegraphics[width=\linewidth]{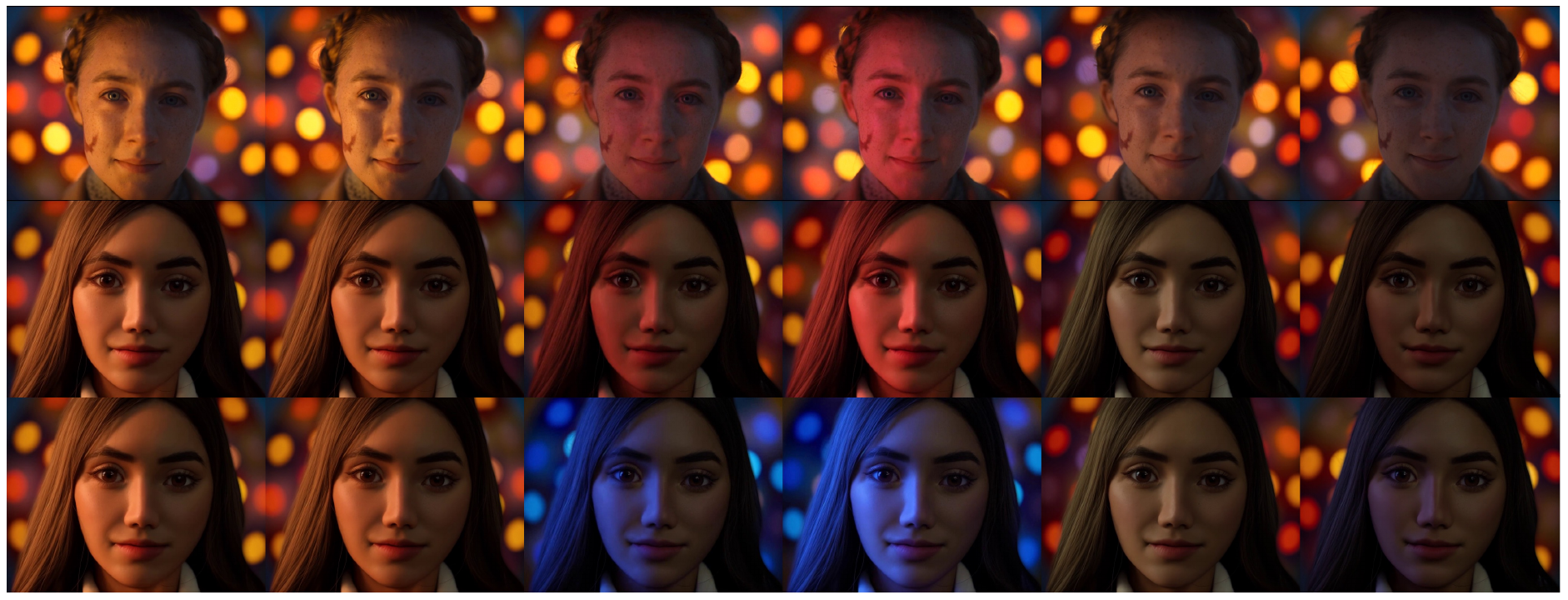}
  \caption{Dynamic Lighting Results. This figure illustrates the replication and editing of dynamic lighting from \textit{The Grand Budapest Hotel} film \cite{budapest}. The first row displays the original scene's dramatic shifts in color and intensity. The second row presents our generated identity under the recreated dynamic lighting, accurately capturing the direction, color, and intensity. In the third row, we modify the original well-designed red lighting to blue, demonstrating our ability to reimagine iconic scenes in new atmospheric conditions while preserving their fundamental essence and adapting to novel lighting scenarios.}
  \label{fig:dynamic_lighting}
\end{figure*}

The essence of Sophia's motion as a virtual actor lies in the digital modulation of motor parameters. This mechanism is central to her ability to precisely control a wide array of facial expressions and head movements. This control enabled Sophia to perform in a way that human actors cannot, with the ability to precisely repeat expressions and actions. To achieve more natural performances, our artists have further refined some keyframes in initial face mocap results using pre-set animation provided by the vendor. This diverse array of pre-set animations encompasses a wide range of emotions and expressions such as joy, frowning, disgust, and blinking. In addition, we use ARKit to further augment the pre-set facial expressions and head movements where the recorded parameters from ARKit are stored and then used on-demand, greatly facilitating reshoots and editing and ensuring that Sophia's performances were both lifelike and consistent. By identifying keyframes that are deemed incorrect or insufficient to reflect the actual expression, the artist chooses the one from the augmented pre-set animations while leaving the rest keyframes untouched. By layering these combinations, we manage to make Sophia more faithfully replicate highly challenging sequences, such as performing the titular song from the feature film \textit{The Sound of Music} \cite{SoundOfMusic} with remarkable consistency.

To better replicate classic movie scenes, we further style Sophia to enrich the versatility as well as for better matching the original clip. These include a range of hairstyles, clothing, and accessories to match the corresponding lighting and performance contents/styles. Examples include clips from feature films like \textit{Scent of a Woman} \cite{scent_of_a_woman}, \textit{The Grand Budapest Hotel} \cite{budapest} and \textit{La La Land} \cite{LaLaLand}. Most of the performance segments by Sophia are around 30 seconds in length, including a range of expressions and actions under different lighting and styling conditions. We also include a special segment where Sophia performs an opera, lasting 2 minutes. This longer performance is significant as it demonstrates the range and capability of Sophia in a more extended format, allowing for a richer display of emotions and a more immersive experience. This variety in performance lengths and styles offers a comprehensive look at Sophia's capabilities as a virtual actor, providing valuable insights into her versatility and potential applications in different cinematic contexts.

\paragraph{Value and Future Potential.}
The SiA Dataset showcases the capability and limitations of the Sophia robot as a virtual performer. We expect it to bring significant stimulation to our community. 
The dataset includes images of Sophia under various cinematic lighting conditions, accompanied by corresponding HDR images, which can be a valuable asset for lighting studies and graphic rendering tasks, such as relighting, environmental light estimation, and inverse rendering. 
The dataset features Sophia in various costumes and dynamic performances, making it ideal for artists and technologists to test virtual shooting techniques and other innovations. 
As we observe tremendous advances in neural modeling and rendering from the neural radiance field (NeRF) \cite{mildenhall2021nerf} to instant-NGP \cite{muller2022instant} and to 3D Gaussian splatting \cite{kerbl3Dgaussians}, the multi-view setup of the SiA dataset may enable new advances that target specifically at virtual performers. 
More importantly, as the first humanoid robotic performance dataset, the SiA dataset is expected to catalyze creative experimentation in filmmaking and digital storytelling as well as offer unique insights to the entertainment industry. 
Currently, Sophia clearly falls short of performing as a human actor, in realism, subtlety, and most importantly, emotion. However, this dataset still showcases the potential benefits of having Sophia in trial shots, as if she is participating in a virtual audition, leading to the title of the paper, Sophia-in-Audition.

%% file: sec/5_idenity.tex
\section{Identity Augmentation on SiA}
\label{sec:identity_augmentation}
Diversity and authenticity are crucial in films for several reasons, reflecting the complexity and richness of the real world and contributing to a more inclusive and representative cinematic landscape.
SiA, aiming to support the film industry, has to address these two challenges in order to fully harness its potential in these creative domains. The first is to produce a more realistic portrayal of Sophia if we want to fully exploit the acquired clips for an audition. Despite her human-like appearance, there are noticeable differences that set her apart from actual humans. These include the way light reflects and scatters off her synthetic skin and the highlights in her eyes. These subtle discrepancies can undermine the realism sought in high-quality film productions.
The second lies in the lack of versatility of Sophia's appearance. Altering her clothes, wigs, or makeup provides some level of diversity, but it falls short of the extensive range often required in character-driven narratives. The idea of possessing different ``Frubber" faces for varied appearances is not economically viable, given the significant resources needed for development and upkeep.

\begin{figure}[ht]
  \centering
  \includegraphics[width=\linewidth]{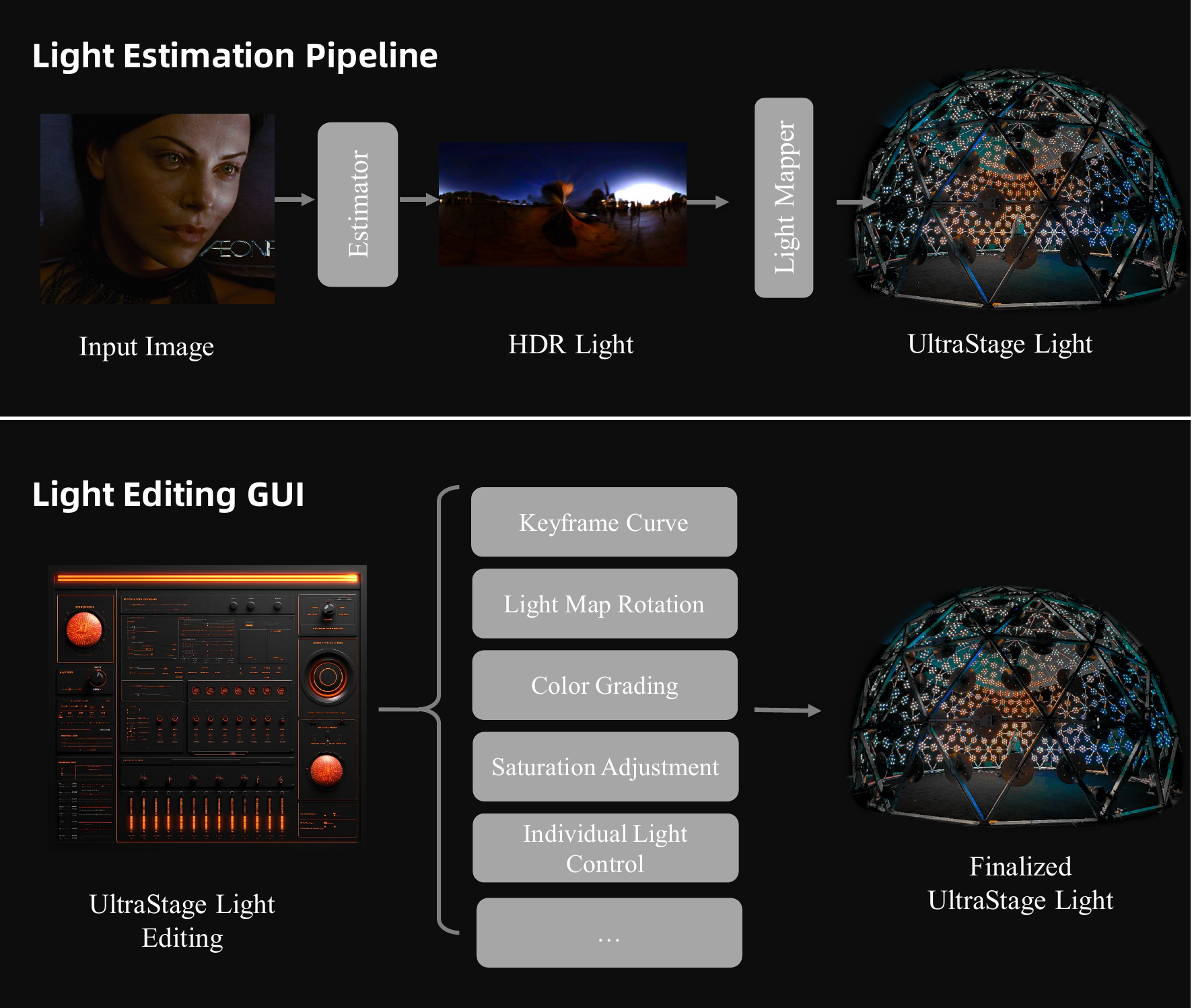}
  \caption{Lighting Simulation Pipeline. We provide a dual track of lighting simulation. The initial track (Fig.\ref{fig:lighting_estimation_pipeline}) begins with an image input, proceeds through a light estimation pipeline to generate an HDR light map, and concludes with a conversion to the 6-channel LED configuration of the UltraStage via the light reproduction process (Fig.~\ref{fig:lighting_reproduction_pipeline}). In complementarity, the second track offers a post-editing graphical user interface (GUI) for the UltraStage, enabling convenient post-processing and refinement of existing light maps. }
  \label{fig:light_edit_pipeline}
\end{figure}

To address these issues, we resort to the latest advancements in the fields of image and video generation. By employing powerful image or video generation models, we show how to digitally augment Sophia's appearance to both enhance her portrayal authenticity/realism and create more diverse versions of her original appearance, allowing for a broader spectrum of character representations than what's physically possible with her current design. Our digital identity augmentation techniques not only circumvent the limitations inherent in Sophia's physical attributes but also align with the evolving digital-centric approaches in film and media production, paving the way for more versatile and realistic applications of humanoid robots like Sophia in the creative industry.

Our specific tasks involve creating highly realistic and temporally coherent neural renderings of a variety of characters, using virtual performer Sophia's performance under various lighting conditions. The main challenge in this task lies in creating characters that are rich in realism and ensuring the temporal stability of the synthesized videos. Latest extensions of stable diffusion (SD) \cite{rombach2022high} based video generation such as AnimateDiff \cite{guo2023animatediff} can generate videos of relatively small motions. In SiA, the head movement and facial expression of Sophia are generally too significant for AnimateDiff. SiA video clips are also much longer than what AnimateDiff can produce (often a few seconds). Our approach aims to address and overcome these limitations to achieve more extended and stable video renderings.

\paragraph{New Character Generation.}
Our general idea is to harness the fundamental capability of SD to generate images from text prompts and images. The goal was to utilize captured performances of Sophia as a medium, enabling SD to create diverse characters while retaining her original expressions, demeanor, poses, and environmental lighting. This was achieved through the img2img method of SD, using Sophia's images as inputs.

The key challenge in directly employing SD for Sophia's appearance enhancement is to balance between generation diversity and temporal consistency. Instead of using SD to generate all the frames, our approach entails a dual-track video generation method that separates the appearance and motion aspects of the process. We start with selecting keyframes from Sophia's performances, ensuring these frames effectively represent the video's local segments. This step is crucial in capturing the essence of her performance. We then apply SD to these keyframes, guided by textual prompts aimed at enhancing and enriching Sophia's identity. This approach allows us to significantly improve her appearance while preserving the original video's major motion, temporal consistency, facial expressions, and environmental lighting. In comparison, direct generation of the entire video clip that maintains high quality across these aspects is extremely challenging whereas our keyframe-based solution is more tractable and very easy to use.

A critical factor in the diffusion process is the control of denoising strength, which fundamentally influences the creative variance of the generated images. We found that lower denoising strength results in images too similar to the input, lacking diversity and quality enhancement. Conversely, higher strength risks significant geometric and textural deviations, leading to inconsistencies across frames. An optimal denoising strength range, determined through experimentation, is between 0.2 and 0.4. To ensure consistency between keyframes and preserve essential elements such as Sophia's facial expressions and the environmental lighting, we further employ ControlNet \cite{controlnet} and its variants such as IP-Adapter \cite{ye2023ip-adapter}. These tools integrate conditions like edge, depth map, facial poses, or identity information into the generation process. The weight of each ControlNet is carefully adjusted, typically between 0.4 and 0.6, based on the specific generation targets and input images. To further enhance image quality and inference speed in SD on SiA clips, we incorporate Latent Consistency Models Low-Rank Adaptation (LCM-LoRA) \cite{luo2023lcmlora}. This technique allows for better-quality images even with fewer sampling steps and lower denoising strength, effectively preserving Sophia's facial expressions and environmental lighting. Utilizing the Euler A sampler with 5-10 sampling steps, a CFG scale between 1.5-3, and ESRGAN\_4X \cite{ESRGAN} as upscaler, we achieved a balance of efficiency and visual fidelity, essential for the realistic portrayal of Sophia in digital media.


To further enhance motion and view consistency across the frames, instead of generating the complete clips, we employ the Example-Based Image Synthesizer (EBSynth) \cite{ebsynth}, which uses an optical-flow-based image matching method. This approach necessitates example frames to guide the image generation process. In our setup, the original video clip served as the motion and consistency reference, while the SD-generated keyframe images were used as appearance references. Selecting the right keyframes was a critical step. We aim to strike a balance between having enough keyframes to maintain visual coherence while avoiding too many, which could degrade temporal stability and introduce artifacts. Recall that the Sophia clips come from facial motion capture where the artists already selected the keyframes. Therefore, these results are recycled here to reduce manual interventions. 

\begin{figure}[ht]
  \centering
  \includegraphics[width=\linewidth]{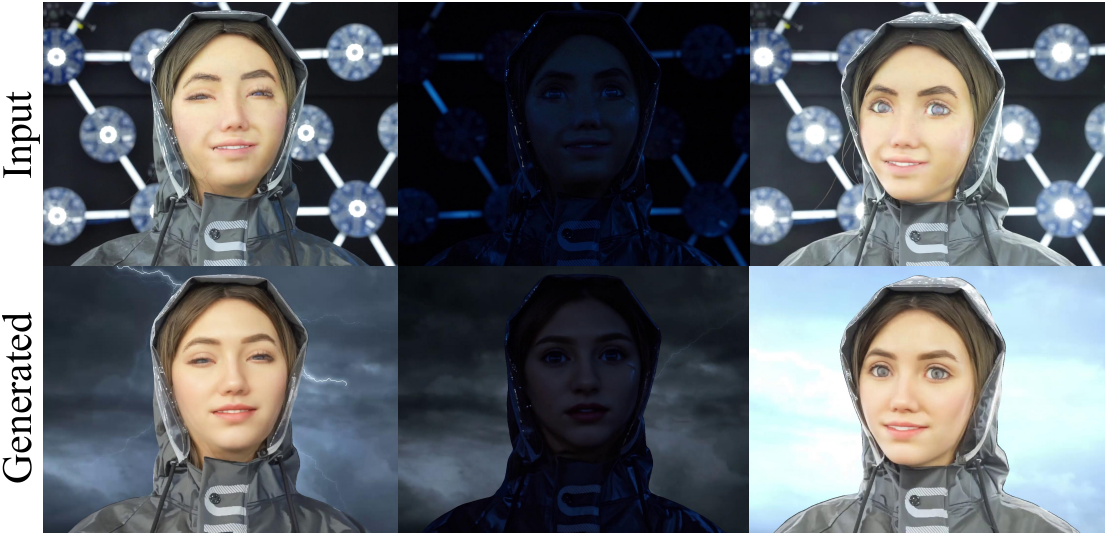}
  \caption{Dynamic Lightning Simulation. UltraStage adeptly simulates the rapid fluctuations characteristic of lightning effects, utilizing dynamic controls for precision. The first row displays three representative frames capturing Sophia within UltraStage, illustrating the initial conditions. Despite the lighting's intense flickering, the second row reveals the generated identity maintains temporal stability and high-quality consistency throughout the sequence.}
  \label{fig:lightning}
\end{figure}

Finally, the geometric consistency of these key frames is paramount in reducing artifacts when applying EBSynth to SiA clips. By starting with these keyframes, we synthesize images, moving either forward or backward to adjacent keyframes. To blend overlapping images from different keyframes, we apply cross-fading that helps greatly in maintaining the continuity and fluidity of the motion while incorporating the enhanced appearance features from the SD-generated keyframes. Figure \ref{fig:identity_aug_result} and the supplementary videos show several examples of our identity-augmented SiA clips. Our Sophia with identity augmentation approach using SD brings unparalleled flexibility in character creation where production teams can quickly generate characters of various appearances and styles based on script requirements or creative inspirations. Overall, we expect our approach will significantly reduce costs and time incurred in casting, makeup, and set design stages, particularly advantageous for independent producers and small studios with limited resources. Furthermore, from an artistic and cultural perspective, SiA brings to life a new unique acting style of unique or surreal characters, such as science fiction or fantasy narratives, e.g., the blockbuster films \textit{M3GAN} \cite{M3GAN} and \textit{Poor Things} \cite{PoorThings}, where traditional casting and makeup might fall short of creative needs.

%% file: sec/6_virutal_lighting_and_cammove.tex
\begin{figure}[ht]
  \centering
  \includegraphics[width=\linewidth]{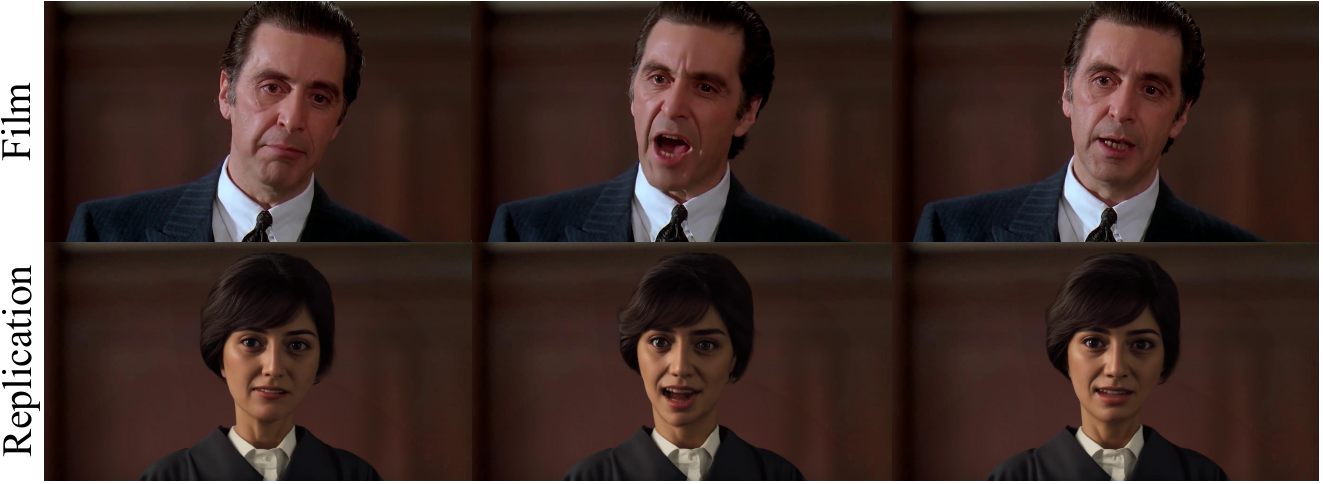}
  \caption{Film Replication Results. Through the estimation and reproduction of lighting from a film and extracting motions in the film via ARKit, we achieved a faithful film recreation. The first row presents the original film clip, while the second row showcases our replicated performance. This comparison illustrates the successful replication of both lighting and facial expressions, adapted to a new identity's appearance (Images courtesy of \textit{Scent of a Woman} \cite{scent_of_a_woman}).}
  \label{fig:replication}
\end{figure}

\begin{figure*}[ht]
  \centering
  \includegraphics[width=\linewidth]{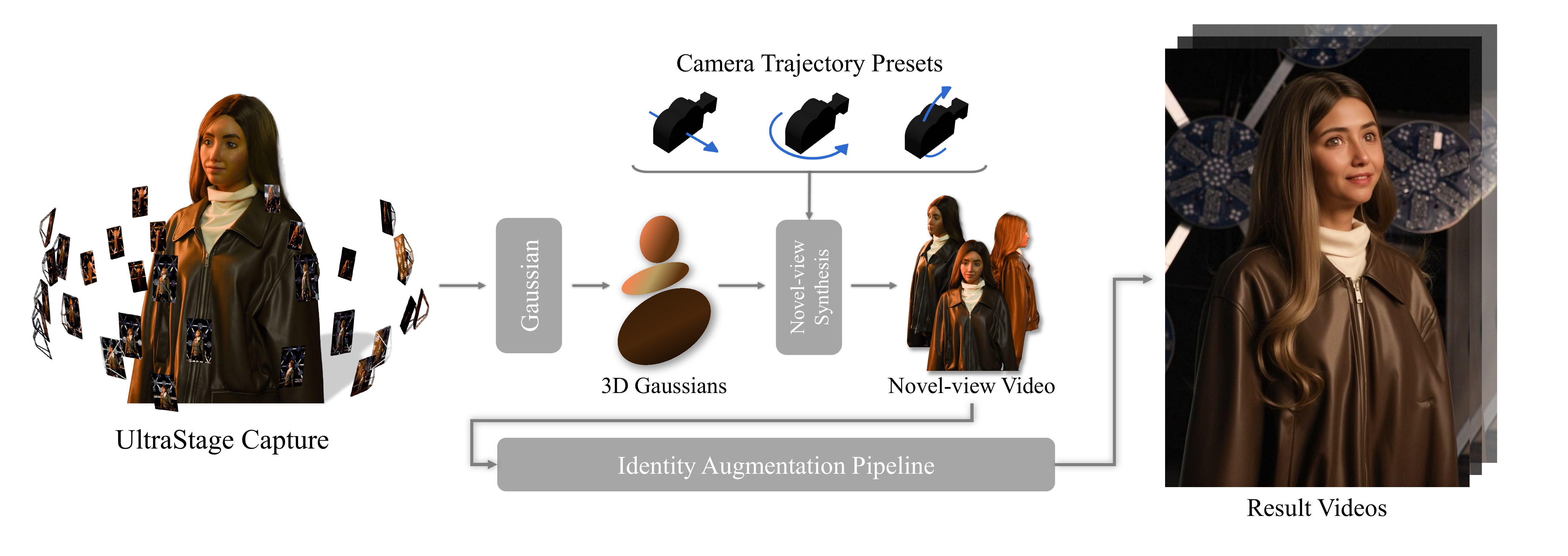}
  \caption{Camera Movement Pipeline. Capturing Sophia's performance from multiple angles in the UltraStage with immersive lighting, SiA enables the application of 3D Gaussian splatting techniques to re-render these performances from new perspectives. This novel-view synthesis incorporates pre-recorded camera trajectories—such as trucking, panning, or tilting—to enhance the visual experience in the synthesized videos. The output videos are then fed into the Identity Augmentation Pipeline(Fig.\ref{fig:identity_aug_pipeline}) for further refinement and quality enhancement.}
  \label{fig:cam_reposition_pipeline}
\end{figure*}

\section{Virtual Lighting and Camera Move}
\label{sec:virtual_lighting_cam_move}
In addition to SiA dataset and video post-processing techniques as discussed in previous sections, we showcase the potential of virtual lighting and camera movement using SiA. Our preliminary studies demonstrate SiA provides an excellent vehicle for filmmakers to experiment with and pre-visualize the complex interplay between lighting, camera angles, and performance, potentially streamlining the pre-production process.

\paragraph{Reproducing Classical Cinematic Lighting.}
Reproducing existing effects is crucial to study their benefits and limitations (same as to SIGGRAPH submissions). We show SiA manages to adeptly resurrect characters within the enchanting glow of classic cinema. Specifically, we set out to recreate iconic film lighting by capturing the authentic essence and visual texture of original movie scenes.

In Figure \ref{fig:light_similation_result}, we present characters adorned in five distinct costumes, illuminated under six types of replicated cinematic lighting setups. Leveraging Stable Diffusion, we achieve more lifelike renderings of "Frubber" skin, showcasing realistic highlights and reflections. Furthermore, the wigs have been accurately adjusted to mirror the nuances of environmental lighting. For example, with top and side lighting, the emulated lighting by UltraStage vividly maintains the integrity of shadows and the dramatic contrast in brightness. The resulting color tones of our replicated lighting convincingly replicate the original cinematic ones, capturing the emotional depth, ambiance, and stylistic nuances of the original film scenes.

A unique capability of SiA is that we can emulate dynamic lighting synchronized with Sophia's performance. 
Building upon prior methods of lighting estimation, we have crafted an interactive graphical user interface that enables users to adjust the color, direction, and intensity of each light panel effortlessly and in real time, as depicted in Figure~\ref{fig:light_edit_pipeline}. Based on this, we can accurately adjust and edit complex environment lighting.
Figure~\ref{fig:dynamic_lighting}, and the supplementary video show the award-winning scene of the Budapest movie. A focal point of the original clip is a dynamic lighting scenario. Here, UltraStage adeptly manages seamless transitions in both color and brightness, ensuring the generated identity's appearance remains consistent throughout the lighting changes. This replication not only highlights UltraStage's ability to produce natural lighting effects but also its capacity to uphold the emotional resonance and visual fidelity of the original cinematic moment. 

More interestingly, we can use SiA to further alter the lighting of a classic movie scene. Such tasks were deemed extremely challenging: it is nearly impossible to re-stage the scene and the lighting condition, let alone have the same actors. We show SiA offers a potential solution. As depicted in the bottom row of Figure~\ref{fig:dynamic_lighting}, UltraStage facilitates inventive lighting modifications, such as altering red lighting to blue during a dynamic sequence. This capability empowers us to view and evaluate iconic scenes within entirely new atmospheric conditions, thereby maintaining their quintessential essence while adapting to innovative lighting environments.
At the same time, it is easy to have Sophia perform the same scene. We also replicate a clip from a famous movie, \textit{Secent of Woman} \cite{scent_of_a_woman}, with corresponding environment lighting and performance. Our user studies (Sec.~\ref{sec:user_study}) show that even though the final results cannot fully match the original clip in quality of performance, they are able to faithfully convey the emotion of orchestrated lighting and performance, sufficiently appealing to the audience and useful to the director.


In Fig.~\ref{fig:lightning}, we further use UltraStage to simulate the thunderstorm lighting. The sudden flashes of lightning and the accompanying thunder create an atmosphere of fear and tension. The unpredictability and intensity of a thunderstorm, coupled with Sophia's performance, contribute to a sense of impending danger or suspense, as shown in the supplementary video. The flickering of lighting is particularly difficult to achieve on previous lighting domes. UltraStage, by employing dynamic controls, manages to produce the correct rhythm as well as temporally coherent visual effects. Achieving such dynamic lighting changes is also challenging using post-capture relighting techniques whereas our workflow manages to handle it with a high precision.

\paragraph{Camera Re-positioning and Virtual Move.}

The significance of camera movement in filmmaking cannot be overstated. Plausible shooting angles and positions greatly support storytelling and encourage audience engagement. In the existing production pipeline, the director will have to choose a specific camera position to first shoot trial shots and then decide whether or not to change the position. A number of pre-production tools attempt to enable the director to preview effects at virtual angles. However, without capturing the real content, it is difficult to assess the quality. There have also been learning-based techniques to conduct post-capture view change \cite{pan2023drag, tewari2020stylerig}. Yet, methods based on 2D-to-3D conversions lack robustness and stability where the results which fall short of cinematic quality. Neural modeling techniques \cite{yu2021pixelnerf, saito2019pifu, feng2021learning, liu2023zero1to3} can partially move around the performer but altering the camera angle at a large magnitude is nearly impossible. A similar but possibly more challenging task is to change the camera angle of a classic clip - it is impractical to conduct reshooting with the same actors from a new desired camera angle. 

\begin{figure*}[t]
  \centering
  \includegraphics[width=\linewidth]{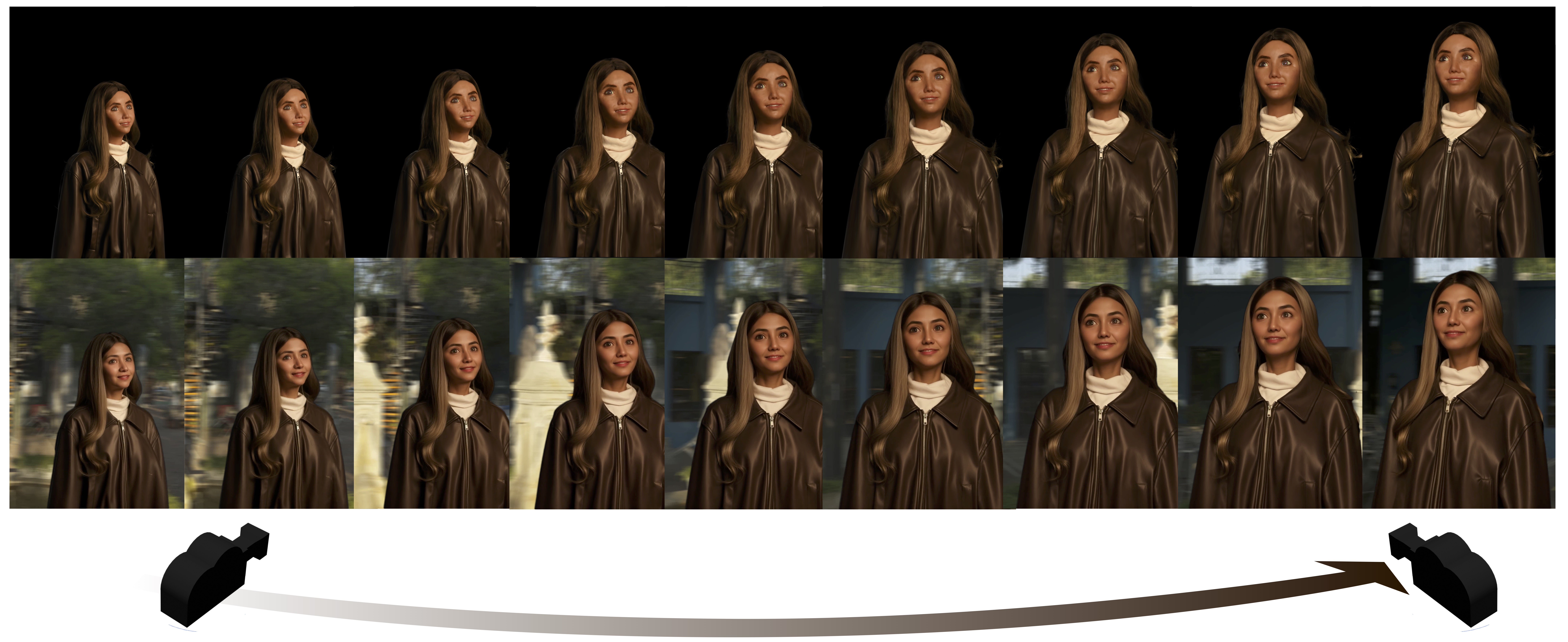}
  \caption{Camera Movement Results. The first row displays the original 3D Gaussian Splatting (3DGS) rendering outcomes, tracing a virtual camera movement path. In the second row, we introduce our generated identity, featuring a significantly enhanced realistic appearance. Remarkably, the visual consistency of this generated identity, especially the butterfly lighting effect on the face, is maintained with high spatial coherence across successive frames, safeguarding uniform visual integrity throughout the sequence.}
  \label{fig:cam_reposition_result}
\end{figure*}

Since UltraStage already incorporates multi-camera capture by design as shown in Fig.~\ref{fig:cam_reposition_pipeline}, this setup can potentially facilitate camera position changes through novel view synthesis. The capture system adopts a sophisticated multi-camera setup, capturing high-resolution 8K images with high-precision synchronization. The cameras are calibrated under ambient white lighting. We use Agisoft Metashape \cite{metashape} to estimate the camera parameters from 32 multi-view images. We use background matting techniques \cite{lin2021real} to extract the foreground from each viewpoint at any given frame, isolating Sophia's performance. These segmented results are subsequently processed through 3D Gaussian splatting (3DGS) for view synthesis and subsequently camera movement and re-positioning. There are a number of solutions that can be potentially deployed for 3D rendering, ranging from the Neural Radiance Fields (NeRF) and its extensions \cite{mildenhall2021nerf, chen2022tensorf, barron2021mip, wang2021neus, yu2021plenoctrees, fridovich2022plenoxels}, in particular, instant-NGP \cite{muller2022instant}, to the latest 3DGS. In our experiments, we find that 3DGS significantly improves rendering quality in scenarios characterized by relatively sparse camera arrangements. We also extend 3DGS to handle multi-view videos of Sophia performing. Instead of processing each multi-view frame individually, we use the same Gaussian spheres of the first frame as the initial for the rest of the frames. This greatly improves efficiency without sacrificing the quality.  

We have also developed a 3DGS-based 3D video rendering to allow camera movement previews. Our renderer employs an interactive viewer that allows for the real-time selection of key camera positions.
Analogous to Luma \cite{luma}, our solution adeptly interpolates between these pivotal points, crafting any desired path for camera movement. The difference though is that we load in 3DGS sequences instead of a single model and provide preview interfaces for temporal domain selection. Our solution allows us to have Sophia replicate performances, where camera positions are altered post-capture, with the results further refined using Stable Diffusion (SD) processing as depicted in Fig.~\ref{fig:cam_reposition_result}. Our ultimate goal is to have SiA seamlessly integrated into the cinematic production workflow, thereby eliminating the need for reshoots due to framing or angle discrepancies.


%% file: sec/7_userstudy.tex
\section{User Studies}
\label{sec:user_study}
Sophia-in-Audition (SiA) is the first of its type that adopts a robot performer in virtual production. It is hence to conduct user studies to assess issues such as whether the results pass the uncanny valley. In our study, we have evaluated multiple aspects of SiA through 116 samples focusing on 4 sample static frames and 2 dynamic video clips under immersive lighting. A more comprehensive study with all clips will be conducted once we disseminate the SiA to the community. Our current study aims to assess two key issues: audience acceptance and affinity towards Sophia's performances, and the effectiveness of immersive environmental lighting in enhancing these performances. The output of the study is instrumental in gauging audience engagement and the realism of Sophia's performance, providing valuable insights for future enhancements in the field of humanoid robotic performers.

In our user study evaluating SiA, we focus on Generation Z (Gen Z), known for growing up in a digital age with widespread access to technology, the internet, and social media. The participant demographic predominantly consists of undergraduate and graduate students aged 18-28 years. These individuals primarily have an academic background in science, engineering, and interaction design. This specific demographic provided a unique perspective, given their familiarity with technological and design principles, thus offering insightful feedback on Sophia's performance and social media oriented contents.


\paragraph{Referential Static Images.} In our user study, we are curious if Sophia's performance in a video surpasses her static synthesized image. Participants rated various aspects of Sophia's performance, in both static image and video, on a scale ranging from -4 to +4. The questions were designed to assess audience acceptance and affinity, focusing on Sophia's appearance, facial expressions, motion, and the psychological responses of the audience. We first presented four static images of Sophia during her performance and asked participants to evaluate the visual quality (-4: Very Rough; +4: Very Polished), attractiveness (-4: Very Unattractive; +4: Very Attractive), and realism (-4: Very Fake; +4: Cannot distinguish between real human) of the character. From 116 samples, these aspects received average scores of +2.53, +1.75, and +1.35, respectively. These results set a high standard for the quality achieved by stable diffusion (SD) and served as a reference scale for comparing with dynamic SiA performance clips.

\paragraph{Motion Smoothness and Naturalness in Videos.}
We further analyzed the participants' responses to video clips of her performance. A particular focus was on motion smoothness and naturalness, with ratings on a scale from -4 to +4(-4: Very Stuttering/Very Unnatural, +4 Very Smooth/Very Natural). The motion smoothness and naturalness scored +0.05 and -0.06, respectively, with about half of the participants (51.73\%, 52.59\%) giving non-negative scores. This suggests that though it still has a long way to go from being considered flawless, the robot's motion is generally accepted as adequate. 

\paragraph{Facial Expressions, Visual Quality, and Attractiveness.} Following with a vote of the diversity and richness of the character's facial expressions (Very deficient, few, moderate, many, Very Expressive), most (75.00\%) of the 116 participants selected non-negative responses, and nearly half (47.41\%) of the participants think Sophia has Many or Very Expressive facial expressions. Also, we asked the same questions as in the static images - the visual quality, attractiveness, realism -with results +0.98, +0.33, -0.47. Compared with static images, +2.53, +1.75, and +1.35, the most significant drop was in realism, which has a 134.81\% decrease. A notable 62.93\% participants (compared to those in static images, only 31.89\% participants) seem to recognize that the video appears less real than images, and they can directly tell that the video clip does not come from a human performer. This is not surprising as Sophia's performance tends to be stiff and mechanic. But more interestingly, the majority (75.00\% and 61.21\%) of 116 participants have a positive view towards Sophia as a performer and find her video clips delicate and attractive.

\paragraph{Viewer Acceptance and Affinity}. We have further explored viewer acceptance and affinity towards Sophia as a virtual performer. Participants were asked about their acceptance of this actor appearing in short videos or films, with responses ranging from -4 (Totally Unacceptable) to +4 (Totally Acceptable). Additionally, we inquired if the actor in the video clips evoked "Uncanny Valley Effects," causing feelings of panic, scare, or discomfort, on a scale from -4 (Very Scared, Panic, or Discomfort) to +4 (Very Enjoyable). The results showed scores of +0.28 and +0.51 for acceptance and affective experience, respectively. Over half of the participants (58.62\% and 64.66\%) indicated a non-negative reaction, finding the virtual performer more pleasurable than troubling. Notably, a small portion (6.9\%) of viewers experienced discomfort. These findings suggest that while the Gen Z audience recognizes the non-human nature of the performer, the majority finds it acceptable and enjoys a positive, affective experience.




\paragraph{Evaluation of Immersive Lighting.} We have further assessed the impact of virtual lighting on the overall quality of performance. Similar to the previous section, we presented the same four static images of SiA during the performance. Viewers were asked to rate the aesthetic appeal of the immersive lighting on a scale from -4 (Not Aesthetic at All) to +4 (Very Aesthetic). Additionally, they evaluated whether the lighting appeared natural on the character's face(-4: Very Unnatural; +4: Very Natural). The responses to these both aspects were positive, with average scores of +1.71 and +1.93, respectively. A significant majority of the participants (87.93\% and 84.49\%) gave non-negative scores, indicating a general agreement with the lighting's aesthetic and natural appearance on the character. We also presented the same video as in the previous section, focusing on assessing the immersive lighting effects. The responses to these questions remain positive, with average scores of +0.88 and +1.01, respectively. A majority of the viewers (68.97\% and 73.28\%) gave non-negative scores, indicating that they generally perceived the virtual lighting in the dynamic SiA clips as aesthetically pleasing and contributing positively to the visual experience. Finally, we have the viewers compare two video clips: the first from \textit{The Grand Budapest Hotel} \cite{budapest} and the second replicating its lighting. Participants rated the lighting's similarity and naturalness of SiA video clip on scales from -4 (Totally Different/Totally Unnatural) to +4 (Very Similar/Very Natural), with the final scores of +1.26 on similarity and +1.09 on naturalness, with 75.85\% and 77.60\% of viewers acknowledging the resemblance and natural appearance between our virtually lit and the original ones, indicating their acceptance of virtual lighting to production.






%% file: sec/8_discussion.tex
\section{Conclusions and Future Directions}
\label{sec:conclusion_future_directions}
Throughout this exploration of the Sophia-in-Audition (SiA) dataset and the accompanying virtual production techniques, we have traversed the intricate landscape of virtual lighting, camera movement, and the replication of iconic cinematic moments. The SiA project, by integrating Sophia into a meticulously controlled UltraStage environment, not only showcases the practical applications of the latest advances in computer graphics and computer vision but also pushes the boundaries of what is achievable in virtual production. Overall, we hope that the SiA project represents a new trial to extend the frontier of entertainment technologies, offering a glimpse into the future of filmmaking. By bridging the gap between traditional cinematography and cutting-edge virtual production methods, SiA provides a robust platform for further development and experimentation. The SiA dataset, the first of its kind, with its comprehensive coverage of lighting scenarios and camera movements, shall serve as a valuable resource for researchers and practitioners alike, stimulating further exploration and new developments in our field.

There are clearly many limitations in this preliminary experiment. Firstly, while Sophia serves as a promising candidate for virtual auditions, her design does not aim to fully replace human actors. Her motor-driven facial expressions, despite being technologically advanced, fall short of capturing the complete subtlety and fluidity of human movements in real time, primarily due to physical limitations. Additionally, the implementation of our dual-track identity augmentation using stable diffusion, although successful in reducing flickering and improving temporal consistency, still presents challenges that detract from the realism of her performances. Moreover, the use of UltraStage, while effective in providing immersive lighting effects, proves to be complex, suggesting the potential for a more streamlined virtual alternative in future developments. Lastly, the user study conducted thus far offers only preliminary insights into SiA's effectiveness. A more exhaustive study encompassing all video clips is planned following the broader dissemination of SiA to the research community, which will provide more comprehensive feedback and data.

Moving forward, it is anticipated that the continued evolution of these technologies will further democratize filmmaking, making high-quality production more accessible and enabling creators to bring their most ambitious visions to life. The SiA dataset, therefore, is not just a testament to the current state of virtual production but also a stepping stone towards the vast, uncharted territories of cinematic storytelling yet to be explored.